\pdfoutput=1
\RequirePackage{ifpdf}
\ifpdf 
\documentclass[pdftex]{sigma}
\else
\documentclass{sigma}
\fi

\numberwithin{equation}{section}
\usepackage{enumerate}
\usepackage[all]{xy}

\newtheorem{Proposition}{Proposition}[section]
{\theoremstyle{definition}
\newtheorem{Remark}[Proposition]{Remark}
}

\usepackage{authblk}

\DeclareMathOperator{\Tr}{Tr}
\DeclareMathOperator{\const}{const}
\DeclareMathOperator{\sign}{sign}
\DeclareMathOperator{\sing}{sing}
\DeclareMathOperator{\reg}{reg}

\begin{document}

\newcommand{\arXivNumber}{1311.0679}

\allowdisplaybreaks

\renewcommand{\PaperNumber}{056}

\FirstPageHeading

\ShortArticleName{Integrable Systems Related to Deformed $\mathfrak{so}(5)$}

\ArticleName{Integrable Systems Related to Deformed $\boldsymbol{\mathfrak{so}(5)}$}

\Author{Alina DOBROGOWSKA and Anatol ODZIJEWICZ}

\AuthorNameForHeading{A.~Dobrogowska and A.~Odzijewicz}

\Address{Institute of Mathematics, University of Bia{\l}ystok, Lipowa 41, 15-424 Bia{\l}ystok, Poland}
\Email{\href{mailto:alaryzko@alpha.uwb.edu.pl}{alaryzko@alpha.uwb.edu.pl},
\href{mailto:aodzijew@uwb.edu.pl}{aodzijew@uwb.edu.pl}}

\ArticleDates{Received November 05, 2013, in f\/inal form May 26, 2014; Published online June 03, 2014}

\Abstract{We investigate a~family of integrable Hamiltonian systems on Lie--Poisson spa\-ces~$\mathcal{L}_+(5)$ dual to
Lie algebras $\mathfrak{so}_{\lambda, \alpha}(5)$ being two-parameter deformations of~$\mathfrak{so}(5)$.
We integrate corresponding Hamiltonian equations on~$\mathcal{L}_+(5)$ and~$T^*\mathbb{R}^5$ by quadratures as well as
discuss their possible physical interpretation.}

\Keywords{integrable Hamiltonian systems; Casimir functions; Lie algebra deformation; symplectic dual pair; momentum map}

\Classification{70H06; 37J15; 53D17}

\section{Introduction}

The notion of compatible Poisson structures on a~manifold~$M$, f\/irstly introduced by Magri in~\cite{5}, leads to one of
the most productive methods of construction of functions on~$M$ being in involution.
This method was used by many authors to integrate various Hamiltonian systems, see, e.g., monograph~\cite{10} for
interesting examples as well as a~huge number of references therein.

A pencil of Lie brackets on vector space $\mathfrak{g}$ def\/ines compatible Lie--Poisson structures on the dual
$\mathfrak{g}^*$ to $\mathfrak{g}$.
For the treatment of this case see~\cite[Chapter 7, Section 44]{10}.
One can f\/ind many examples of Hamiltonian systems on Lie--Poisson space $\mathfrak{g}^*$ obtained in this way
in~\cite{27aa1,27aa,27aa5,5,5aa,14aa, 27aa4, 27aa3, 27aa2}.

In~\cite[Section~3, Proposition~6]{7} we investigate compatible Lie--Poisson structures on space~$\mathcal{L}_+$ of
uppertriangular Hilbert--Schmidt operators.
Since this case includes all f\/inite-dimensional cases~$\mathcal{L}_+(n)$, $n\in\mathbb{N}$, we will come to f\/inite-dimensional integrable Hamiltonian systems related to various Lie algebras whose Lie brackets depends on a~f\/inite number
of real parameters.
Within this context in the present paper we consider a~two-parameter family of Lie algebras $\mathfrak{so}_{\lambda,
\alpha}(5)$, $\lambda, \alpha\in \mathbb{R}$, which contains physically important subcases such as Poincar\'e algebra,
Galilean algebra, de Sitter algebra, anti-de Sitter algebra, special orthogonal algebra $\mathfrak{so}(5)$ and
Euclidean algebra $\mathfrak{e}(4)$.
We arrange all these cases in the table below
\begin{center}
\begin{tabular}{|c|c|c|}
\hline
1) & $\lambda>0
\quad
\wedge
\quad
\alpha >0$ & $\mathfrak{so}(5)$
\\
\hline
2) & $\lambda <0
\quad
\wedge
\quad
\alpha >0$ & $\mathfrak{so}(3,2)\simeq \mathfrak{sp}(2,\mathbb{R})$
\\
\hline
3) & $\lambda<0
\quad
\wedge
\quad
\alpha <0$ & $\mathfrak{so}(1,4)$
\\
\hline
4) & $\lambda<0
\quad
\wedge
\quad
\alpha =0$ & $\mathfrak{p}(1,3)$ (Poincar\'e algebra)
\\
\hline
5) & $\lambda=0
\quad
\wedge
\quad
\alpha=0$ & Galilean algebra
\\
\hline
6) & $\lambda>0
\quad
\wedge
\quad
\alpha =0$ & $\mathfrak{e}(4)$ (Euclidean algebra)
\\
\hline
7) & $\lambda=0
\quad
\wedge
\quad
\alpha> 0$ & $\left(\mathfrak{so}(2) \times \mathfrak{so}(3) \right)\ltimes {\rm Mat}_{3\times 2}(\mathbb{R})$
\\
\hline
8) & $\lambda=0
\quad
\wedge
\quad
\alpha< 0$ & $\left(\mathfrak{so}(1,1) \times \mathfrak{so}(3) \right)\ltimes {\rm Mat}_{3\times 2}(\mathbb{R})$
\\
\hline
\end{tabular}
\end{center}

The physical importance of the above Lie algebras motived us to investigate related Hamiltonian systems.
The Hamiltonian systems connected with Euclidean, Galilean and Poincar\'e Lie algebras, specif\/ied by the condition
$\alpha=0$, were studied (integrated) in~\cite{3}.
We will study here other cases, characterized by the condition $\alpha\lambda\neq 0$, i.e.~the ones corresponding to
$\mathfrak{so}(5)$, $\mathfrak{so}(1,4)$ and $\mathfrak{so}(3,2)$.
Let us mention that the construction of integrals of motion in involution on $\mathcal{L}_+(n)$ proposed in~\cite{7} for
$n=4$ leads to linear Hamiltonian systems.
The Hamiltonian systems obtained for the case $n>5$ depend on more deformation parameters and thus are more dif\/f\/icult to be handled.

The main results of the paper are the following ones.
In Section~\ref{Section2} we construct and integrate by quadratures a~Hamiltonian system on Lie--Poisson space $\mathcal{L}_+(5)$
with Poisson bracket $\{\cdot,\cdot \}_{\lambda,\alpha}$ def\/ined by~\eqref{6} and Hamiltonian def\/ined by~\eqref{a51}.

In Section~\ref{Section3} we f\/ind the momentum map ${\cal{J}}: T^{*}\mathbb{R}^5\rightarrow \mathcal{L}_+(5) \cong \mathfrak{so}(5)$
of the cotangent bund\-le~$T^*\mathbb{R}^5$ into Lie--Poisson space $\mathcal{L}_+(5)$.
Then $T^*\mathbb{R}^5\setminus {\cal J}^{-1}(0)$ is shown to be the total space of ${\rm GL}(2,\mathbb{R})$-principal bundle
over Grassmannian $G(2,5)$.
We also def\/ine other momentum map $ {\cal{I}}: T^{*}\mathbb{R}^5\rightarrow \mathfrak{sl}(2,\mathbb{R}) \simeq
\mathfrak{sl}(2,\mathbb{R})^*$ and show that $T^*\mathbb{R}^5$ and Lie--Poisson spaces $\mathfrak{sl}(2,\mathbb{R})$
and $\mathfrak{so}(5)$ form symplectic dual pair in the sense of def\/inition presented in~\cite[Chapter IV, Section 9.3]{11}.
Further the spliting of ${\cal J}(T^*\mathbb{R}^5\setminus {\cal J}^{-1}(0))$ on co-adjoint
${\rm SO}_{\lambda,\alpha}$-orbits is given.

The lifting of the Hamiltonian system~\eqref{a52}--\eqref{a52a3} on the symplectic manifold $T^*\mathbb{R}^5$, see
Hamiltonian~\eqref{e4} and Hamilton equations~\eqref{e7}, is integrated in Section~\ref{Section4}.
We present some examples of the physical interpretation of the system given by~\eqref{e4} in Section~\ref{Section4} as well.

\section[Compatible Poisson structures related to deformed $\mathfrak{so}_{\lambda,\alpha}(5)$]{Compatible Poisson structures related to deformed $\boldsymbol{\mathfrak{so}_{\lambda,\alpha}(5)}$}
\label{Section2}

By def\/inition two Poisson brackets $\{\cdot, \cdot\}_1$ and $\{\cdot, \cdot\}_2$ on a~manifold~$M$ are compatible if any
linear combination $b_1\{\cdot, \cdot\}_1 + b_2 \{\cdot, \cdot\}_2$ is also a~Poisson bracket.
If $\{\cdot, \cdot\}_2$ is not a~scalar multiple of $\{\cdot, \cdot\}_1$ then using well elaborated methods, e.g.\
see~\cite{27,5,10}, one can construct integrable Hamiltonian systems on~$M$.
In this section, basing on the paper~\cite{7}, we def\/ine such systems in the case when~$M$ is the vector space of
strictly uppertriangular $5\times 5$~matrices $\mathcal{L}_+(5)$.

Let us consider the vector space $\mathfrak{so}_{\lambda,\alpha}(5)$ of matrices
\begin{gather*}
{\boldsymbol X}:= \left(
\begin{matrix}
0 & \alpha b & \alpha \lambda \vec{u}^{\top}
\\
-b & 0 &\lambda \vec{w}^{\top}
\\
-\vec{u} & -\vec{w} & \delta
\end{matrix}
\right)\in {\rm Mat}_{5\times 5}(\mathbb{R})
\end{gather*}
with f\/ixed parameters $\alpha,\lambda \in \mathbb{R}$ and $b \in \mathbb{R}$, $\vec{u}, \vec{w}\in \mathbb{R}^{3}$,
$\delta \in \mathfrak{so}(3)$.
One easily verif\/ies that $\mathfrak{so}_{\lambda,\alpha}(5)$ is a~Lie algebra with respect to the standard matrix
commutator.

Using the pairing
\begin{gather}
\label{5}
\langle  {\boldsymbol X}, \kappa \rangle :=\operatorname{Tr} (\kappa {\boldsymbol X}),
\end{gather}
between ${\boldsymbol X}\in \mathfrak{so}_{\lambda, \alpha}(5)$ and
\begin{gather*}
\kappa =\left(
\begin{array}{@{}cc|c@{}} 0 & a~& \vec{x}^{\top}
\\
0 & 0 & \vec{y}^{\top}
\\
\hline
\vec{0} & \vec{0} & {\boldsymbol \mu}\tsep{1pt}
\end{array}
\right)\in \mathcal{L}_+(5),
\end{gather*}
where $a\in\mathbb{R}$, $\vec{x},\vec{y}\in \mathbb{R}^{3}$ and
\begin{gather*}
{\boldsymbol \mu}=\left(
\begin{matrix}
0 & \mu_3 & -\mu_2
\\
0 & 0 & \mu_1
\\
0 & 0 & 0
\end{matrix}
\right) \in {\rm Mat}_{3\times 3}\left(\mathbb{R}\right),
\end{gather*}
we will identify $\mathcal{L}_+(5)$ with the dual $\mathfrak{so}_{\lambda, \alpha}(5)^*$ of $\mathfrak{so}_{\lambda, \alpha}(5)$.

In the coordinates $(a,\vec{x},\vec{y},{\boldsymbol \mu})$ Lie--Poisson bracket for $f,g\in C^{\infty}({\cal L}_+(5))$
is given by the formula
\begin{gather}
\{f,g\}_{\lambda, \alpha}= \operatorname{Tr}\left(\kappa \left[\frac{\partial f}{\partial \kappa},\frac{\partial
g}{\partial \kappa}\right]\right)= \lambda a\bigg(\frac{\partial f}{\partial\vec{x}}\cdot \frac{\partial g}{\partial
\vec{y}} -\frac{\partial f}{\partial \vec{y}}\cdot \frac{\partial g}{\partial \vec{x}}\bigg)
\nonumber
\\
\phantom{\{f,g\}_{\lambda, \alpha}=}
{}+ \vec{\mu}\cdot\bigg(\alpha\lambda\left(\frac{\partial f}{\partial\vec{x}}\times \frac{\partial g}{\partial
\vec{x}}\right) +\lambda\left(\frac{\partial f}{\partial \vec{y}}\times \frac{\partial g}{\partial
\vec{y}}\right)+\left(\frac{\partial f}{\partial\vec{\mu}}\times \frac{\partial g}{\partial\vec{\mu}}\right) \bigg)
\nonumber
\\
\phantom{\{f,g\}_{\lambda, \alpha}=}
{}+ \frac{\partial g}{\partial a}\vec{x}\cdot \frac{\partial f}{\partial \vec{y}} -\frac{\partial f}{\partial
a}\vec{x}\cdot \frac{\partial g}{\partial \vec{y}}-\alpha \frac{\partial g}{\partial a}\vec{y}\cdot \frac{\partial
f}{\partial \vec{x}}+\alpha \frac{\partial f}{\partial a}\vec{y}\cdot \frac{\partial g}{\partial \vec{x}}
\nonumber
\\
\phantom{\{f,g\}_{\lambda, \alpha}=}
{}+ \vec{x}\cdot\bigg(\frac{\partial f}{\partial\vec{x}}\times \frac{\partial g}{\partial \vec{\mu}} +\frac{\partial
f}{\partial \vec{\mu}}\times \frac{\partial g}{\partial \vec{x}}\bigg)+ \vec{y}\cdot\bigg(\frac{\partial
f}{\partial\vec{y}}\times \frac{\partial g}{\partial \vec{\mu}} +\frac{\partial f}{\partial \vec{\mu}}\times
\frac{\partial g}{\partial \vec{y}}\bigg),
\label{6}
\end{gather}
where $\vec{\mu}= (\mu_1, \mu_2,\mu_3 )^{\top}$.
Let us note that this bracket belongs to the family of Lie--Poisson brackets investigated in~\cite{7}.

According to Proposition 3 from~\cite{7} the global Casimirs for the bracket $\{\cdot,\cdot\}_{\lambda, \alpha}$ are as
follows
\begin{gather}
\label{ab55}
c_1=\vec{x}\,{}^2+\alpha\vec{y}\,{}^2+\alpha\lambda \vec{\mu}^2+\lambda a^2,
\\
\label{ab56}
c_2=\alpha\lambda (\vec{\mu}\cdot \vec{y})^2 +\lambda(\vec{\mu}\cdot \vec{x})^2+ (\lambda a\vec{\mu}-\vec{x}\times\vec{y} )^2.
\end{gather}
Choosing a~Hamiltonian $H\in C^{\infty}({\cal L}_+(5))$ we obtain Hamilton equations
\begin{gather}
\label{8}
\frac{da}{dt}=\alpha\vec{y}\cdot\frac{\partial H}{\partial \vec{x}} -\vec{x}\cdot\frac{\partial H}{\partial \vec{y}},
\\
\label{9}
\frac{d\vec{x}}{dt}=-\alpha\frac{\partial H}{\partial a}\vec{y}+ \alpha\lambda\frac{\partial H}{\partial
\vec{x}}\times\vec{\mu} +\lambda a\frac{\partial H}{\partial \vec{y}}+ \frac{\partial H}{\partial \vec{\mu}}\times
\vec{x},
\\
\label{10}
  \frac{d\vec{y}}{dt}=\frac{\partial H}{\partial a}\vec{x}+ \lambda\frac{\partial H}{\partial \vec{y}}\times\vec{\mu}
-\lambda a\frac{\partial H}{\partial \vec{x}}+ \frac{\partial H}{\partial \vec{\mu}}\times \vec{y},
\\
\label{11}
  \frac{d\vec{\mu}}{dt}=- \vec{x}\times\frac{\partial H}{\partial \vec{x}}- \vec{y}\times\frac{\partial H}{\partial
\vec{y}}- \vec{\mu}\times\frac{\partial H}{\partial \vec{\mu}}
\end{gather}
on Lie--Poisson space ${\cal L}_+(5)$.
We will construct a~family of Hamiltonians depending on two real parameters, which are completely integrable.

To this end we observe that Poisson brackets $\{\cdot,\cdot \}_{\lambda,\alpha}$ and $\{\cdot,\cdot \}_{\epsilon,\beta}$
are compatible if $\alpha=\beta$ or $\lambda=\epsilon$, see~\cite[Proposition 4]{7} what means that the linear
combination of these brackets is a~Lie--Poisson bracket.
In this paper we consider the case when $\lambda\neq \epsilon$ and $\alpha=\beta$.
Since the case $\alpha= 0$ was considered in~\cite{3} we will not discuss it here.
The bi-Hamiltonian systems given by the Lie--Poisson bracket $\{\cdot,\cdot \}_{1,\alpha}$ and the constant
Lie--Poisson bracket were studied in~\cite{20}.

By Magri method~\cite{5} it can be shown that Casimir functions of the Poisson bracket
$\{\cdot,\cdot\}_{\epsilon,\alpha}$:
\begin{gather}
\label{a55}
h_1=\vec{x}\,{}^2+\alpha\vec{y}\,{}^2+\alpha\epsilon\vec{\mu}^2+\epsilon a^2,
\\
\label{a56}
h_2=\alpha \epsilon(\vec{\mu}\cdot \vec{y})^2 +\epsilon(\vec{\mu}\cdot \vec{x})^2+ (\epsilon
a\vec{\mu}-\vec{x}\times\vec{y} )^2.
\end{gather}
are in involution with respect to the Poisson bracket $\{\cdot,\cdot\}_{\lambda,\alpha}$.

For Hamiltonian
\begin{gather}\label{a51}
H=\gamma h_1+\nu h_2,
\end{gather}
where $\gamma, \nu\in \mathbb{R}$, equations~\eqref{8}--\eqref{11} take the form
\begin{gather}
\label{a52}
\frac{d a}{dt}= 0,
\\
\label{a52a1}
\frac{d\vec{\mu}}{dt}=0,
\\
\frac{d\vec{x}}{dt}=  2(\lambda-\epsilon)\big(\gamma\alpha(a\vec{y}+\vec{x}\times\vec{\mu})+
\nu(\alpha\vec{\mu}\times((\vec{x} \times\vec{y})\times\vec{y})
\nonumber
\\
\phantom{\frac{d\vec{x}}{dt}=}
{}+\alpha\epsilon a\vec{\mu}^2\vec{y}+ \epsilon a^2\vec{x}\times\vec{\mu} + a(\vec{x}\times \vec{y})\times\vec{x})\big),
\label{a52a2}
\\
\frac{d\vec{y}}{dt}=  2(\lambda-\epsilon)\big(\gamma(-a\vec{x}+\alpha \vec{y}\times\vec{\mu})+
\nu(\vec{\mu}\times ((\vec{y}\times\vec{x})\times\vec{x})
\nonumber
\\
\phantom{\frac{d\vec{y}}{dt}=}
{}-\epsilon a\vec{\mu}^2\vec{x}+\epsilon a^2\vec{y}\times\vec{\mu}+ a(\vec{x}\times \vec{y})\times \vec{y})\big).
\label{a52a3}
\end{gather}

One can verify functions $h_1, h_2, \vec{\mu}^2, \mu_3, a, h_1-c_1, h_2-c_2\in C^{\infty}({\cal L}_+(5))$ to be
integrals of motion here which are in involution.
Recall that $c_1$ and $c_2$ are Casimir functions def\/ined in~\eqref{ab55},~\eqref{ab56}.
Since generic symplectic leaves of $\mathcal{L}_+(5)$ have dimension eight then for the integrability of the above
Hamiltonian system it is enough to possess four functionally independent integrals of motion being in involution with
respect to the Poisson bracket $\{\cdot,\cdot\}_{\lambda,\alpha}$.
For example one of the possible choices of four integrals of motion is
\begin{gather}
I_1:=a,
\qquad
I_2:=\mu_3,
\qquad
I_3:=h_1-c_1=(\epsilon -\lambda)\big(\alpha \vec{\mu}^2 + a^2\big),
\nonumber
\\
I_4:=h_2-c_2= \alpha (\epsilon -\lambda) (\vec{\mu}\cdot \vec{y} )^2 +(\epsilon
-\lambda) (\vec{\mu}\cdot \vec{x} )^2
\nonumber
\\
\phantom{I_4:=}
{}+\big(\epsilon^2 -\lambda^2\big)a^2\vec{\mu}^2-2(\epsilon -\lambda) a\vec{\mu}\cdot (\vec{x} \times \vec{y} ).
\label{l1c}
\end{gather}
One easily verif\/ies that the following proposition is valid.
\begin{Proposition}
The Jacobi matrix $DI(a,\vec{\mu},\vec{x},\vec{y})$ of the map $I:\mathcal{L}_+(5)\rightarrow \mathbb{R}^4$ defined
by~\eqref{l1c} has rank smaller than four if and only if
\begin{gather}
\label{l2c}
\mu_i\left( (\vec{\mu}\cdot\vec{x} )\vec{\mu}+a\vec{\mu}\times \vec{y}\right)=0
\quad
\wedge
\quad
\mu_i\left(\alpha  (\vec{\mu}\cdot\vec{y} ) \vec{\mu} -a\vec{\mu}\times \vec{x}\right)=0,\qquad i=1,2.
\end{gather}
\end{Proposition}

From~\eqref{l2c} we conclude that $a$, $\mu_3$, $h_1-c_1$, $h_2-c_2$ are integrals of motion functionally independent almost everywhere.
There are the other choices of four integrals of motion for example $a$, $\vec{\mu}^2$, $h_1$, $h_2$, which are also
functionally independent almost everywhere.
However, the proof of this property is technically more dif\/f\/icult than in the case~\eqref{l1c}.

Now we integrate the Hamiltonian equations~\eqref{a52}--\eqref{a52a3} by quadratures.
For this reason we mention that $\vec{\mu}$ and~$a$ are integrals of motion.
Hamiltonian~\eqref{a51} is invariant with respect to the action of the rotation group $ {\rm SO}(3)$ def\/ined by
\begin{gather*}
(a,\vec{x},\vec{y},\vec{\mu})\rightarrow (a, O\vec{x},O\vec{y}, O\vec{\mu}),
\end{gather*}
where $O\in {\rm SO}(3)$.
The above motivates us to use the following ${\rm SO}(3)$-invariant coordinates
\begin{gather}
x:=\vec{\mu}\cdot \vec{x},
\qquad
y:=\vec{\mu}\cdot \vec{y},
\qquad
f:=2\vec{x}\cdot \vec{y},
\label{a58}
\end{gather}
in order to solve~\eqref{a52}--\eqref{a52a3}.
In these coordinates equations~\eqref{a52a2},~\eqref{a52a3} (for the case $\alpha\neq 0$, $a\neq 0$) reduce to the
following three equations
\begin{gather}
\label{a60}
\frac{d}{dt}\left(
\begin{matrix}
x
\\
y
\end{matrix}
\right) =(\lambda-\epsilon)\nu a\left(
\begin{matrix}
-f & \alpha K\pm \sqrt{\alpha}\sqrt{C-f^2}
\\
-K\pm \frac{1}{\sqrt{\alpha}}\sqrt{C-f^2} & f
\end{matrix}
\right) \left(
\begin{matrix}
x
\\
y
\end{matrix}
\right),
\\
\label{a59}
\frac{df}{dt}=\pm 2(\lambda-\epsilon)\alpha\nu a\sqrt{C-f^2}\left(\frac{1}{a^2}\big(x^2+\alpha y^2\big)+D\right),
\end{gather}
where the constants~$C$,~$D$ and~$K$ are expressed is terms of Casimirs~\eqref{ab55},~\eqref{ab56} and integrals of
motion~$a$, $h_1$, $h_2$ and $\vec{\mu}^2$ in the following way
\begin{gather*}
C=\alpha^{-1}\big(c_1-\alpha\lambda\vec{\mu}^2-\lambda a^2\big)^2-4\left(\frac{\lambda h_2-\epsilon
c_2}{\lambda-\epsilon}+\lambda\epsilon a^2\vec{\mu}^2\right),
\\
D=\frac{h_2-c_2}{a^2(\lambda-\epsilon)}-2\epsilon\vec{\mu}^2-\frac{2\gamma}{\nu},
\qquad
K=\alpha^{-1}\big(c_1-\alpha\lambda \vec{\mu}^2-\lambda a^2\big)+ 2\epsilon \vec{\mu}^2+\frac{2\gamma}{\nu}.
\end{gather*}

Introducing new variables~$\varphi$,~$\psi$ and~$r$ by
\begin{gather*}
f:=\sqrt{C}\cos \varphi,
\qquad
x:=e^r\sqrt{\alpha}\cos \left(\frac 12\left(\psi\mp \varphi\right)\right),
\qquad
y:=e^r\sin \left(\frac 12\left(\psi\mp \varphi\right)\right)
\end{gather*}
and substituting them into~\eqref{a60} and~\eqref{a59} we obtain
\begin{gather}
\label{a64}
\frac{dr}{dt}=-\nu a\sqrt{C}(\lambda-\epsilon)\cos \psi,
\\
\label{a65}
\frac{d \psi}{dt}=2\nu a\sqrt{C}(\lambda-\epsilon) \left(\sin \psi-K\sqrt{\frac{\alpha}{C}} -
\left(\frac{\alpha^2}{a^2\sqrt{C}}e^{2r}+\frac{D\alpha}{\sqrt{C}}\right)\right),
\\
\label{a66}
\frac{d\varphi}{dt}=\mp 2\alpha\nu a(\lambda-\epsilon) \left(\frac{\alpha}{a^2}e^{2r}+D\right).
\end{gather}
Now from~\eqref{a64} and~\eqref{a65} we have
\begin{gather}
\label{a67}
\frac 12 \frac{g'(t)}{1-g^2(t)}+Eg(t)-4\nu^2 a^2C(\lambda-\epsilon)^2g^2(t)=:R=\const,
\end{gather}
where
\begin{gather*}
  g(t):=\sin \psi(t),
\qquad
  E=8\nu^2 a^2\sqrt{C}(\lambda-\epsilon)^2(K\sqrt{\alpha} \pm D\alpha).
\end{gather*}
Separating variables in~\eqref{a67} we f\/ind
\begin{gather*}
t=\int \frac{dg}{\sqrt{(g^2-1)(Eg-4\nu^2 a^2C(\lambda-\epsilon)^2g^2-R)}},
\end{gather*}
where constant~$R$ is def\/ined by~\eqref{a67}.
Functions $x(t)$, $y(t)$, $f(t)$ are expressed by means of elliptic function $g(t)$ as follows
\begin{gather*}
f(t)=\sqrt{C}\cos \bigg(\mp 2(\lambda-\epsilon)\alpha\nu a\bigg(D(t-t_0)
+ \frac{\alpha}{a^2}\int_{t_0}^{t}e^{-2\nu a\sqrt{C}(\lambda-\epsilon) \int_{s_0}^{s}\sqrt{1-g^2(z)}dz}ds
\bigg)\bigg),
\\
x(t)=\sqrt{\alpha} e^{-\nu a\sqrt{C}(\lambda-\epsilon)\int_{t_0}^{t}\sqrt{1-g^2(s)}ds}\cos
\bigg((\lambda-\epsilon)\nu a\sqrt{C}\bigg(\int_{t_0}^{t}g(s)ds-K\sqrt{\frac{\alpha}{C}}(t-t_0)\bigg)\bigg),
\\
y(t)=e^{-\nu a\sqrt{C}(\lambda-\epsilon)\int_{t_0}^{t}\sqrt{1-g^2(s)}ds}\sin\bigg((\lambda-\epsilon)\nu
a~\sqrt{C}\bigg(\int_{t_0}^{t}g(s)ds-K\sqrt{\frac{\alpha}{C}}(t-t_0)\bigg)\bigg).
\end{gather*}

Now, without loss of generality, we can assume $\vec{\mu}=(0,0,\mu)$.
Then we obtain that
\begin{gather}
x_3(t)=\frac{1}{\mu}x(t),
\qquad
y_3(t)=\frac{1}{\mu}y(t).
\label{m5}
\end{gather}
One obtains the other coordinate functions $x_1(t)$, $x_2(t)$, $y_1(t)$ and $y_2(t)$ from algebraic equations{\samepage
\begin{gather}
\frac 12f(t)-\frac{1}{\mu^2}x(t)y(t) = x_1(t)y_1(t)+x_2(t)y_2(t),
\nonumber
\\
h_1-\alpha\epsilon \mu^2-\epsilon a^2-\frac{1}{\mu^2}\big(x^2(t)+\alpha y^2(t)\big) = x_1^2(t)+x_2^2(t)+\alpha y_1^2(t)+\alpha y_2^2(t),
\nonumber
\\
\left(x_1^2(t)+x_2^2(t)+\frac{1}{\mu^2}x^2(t)\right) \left(y_1^2(t)+y_2^2(t)+\frac{1}{\mu^2}y^2(t)\right)
=\frac{\epsilon c_2-\lambda h_2}{\epsilon-\lambda}+\epsilon\lambda a^2\mu^2+\frac 14 f^2(t),
\nonumber
\\
\label{m6}
 \frac{c_2-h_2}{\lambda-\epsilon}-\alpha y^2(t)-x^2(t)-(\epsilon+\lambda)a^2\mu^2=2a\mu\left(x_2(t)y_1(t)-x_1(t)y_2(t)\right)
\end{gather}
which follow from~\eqref{ab56}~\eqref{a55},~\eqref{a56},~\eqref{a58}.}

In order to integrate equations~\eqref{a52}--\eqref{a52a3} in the case when $a=0$ we note that functions
\begin{gather}
g_1:=\vec{\mu} \cdot \vec{x}(t),
\qquad
g_2:=\vec{\mu} \cdot \vec{y}(t),
\qquad
g_3:=\left(\vec{x}(t) \times \vec{y}(t)\right)^2=c_2-\lambda g_1-\alpha\lambda g_2
\label{l5c}
\end{gather}
are independent of the parameter $t\in\mathbb{R}$.
We note also that functions
\begin{gather}
f_1:=\vec{\mu}\cdot \left(\vec{x}\times \vec{y}\right),
\qquad
f_2:=\vec{x}\cdot \vec{y},
\qquad
f_3:=\vec{x}\,{}^2-\alpha\vec{y}\,{}^2,
\label{ll6cc}
\end{gather}
satisfy system of equations
\begin{gather}
\frac{d}{dt}f_1=2(\lambda-\epsilon)\nu\big(\alpha g_2^2-g_1^2\big) f_2+ 2 (\lambda-\epsilon)\nu g_1g_2 f_3,
\nonumber
\\
\label{l4c}
\frac{d}{dt}f_2= 2 (\lambda-\epsilon)\nu f_1f_3,
\\
\frac{d}{dt}f_3= -8 (\lambda-\epsilon)\alpha\nu f_1f_2.
\nonumber
\end{gather}
From~\eqref{l4c} we f\/ind
\begin{gather*}
4\alpha f_2^2+f_3^2=M=\const,
\qquad
f_1^2+\frac{1}{4\alpha}\big(\alpha g_2^2-g_1^2\big)f_3-2 g_1g_2 f_2=N=\const,
\end{gather*}
and thus the equation
\begin{gather*}
\frac{d}{dt}f_3= \pm 2 (\lambda-\epsilon)\alpha \nu \sqrt{\alpha^{-1}\big(M-f_3^2\big) \left(4N-
\alpha^{-1}\big(\alpha g_2^2-g_1^2\big) f_3\pm 4g_1g_2\sqrt{\alpha^{-1}\big(M-f_3^2\big)}\right)}
\end{gather*}
holds.
This equation is solved by quadratures.
Finally we f\/ind $\vec{x}(t)$ and $\vec{y}(t)$ solving the algebraic system of equation given by~\eqref{l5c},~\eqref{ll6cc}.

\section{Symplectic dual pair}\label{Section3}

In this section we will consider the case $\alpha\lambda\neq 0$.
Using Pl\"{u}cker embedding we will def\/ine momentum map ${\cal J}: T^{*}\mathbb{R}^5\rightarrow {\cal L_+}(5)\cong
\mathfrak{so}(5)$ for the canonical action of ${\rm SO}_{\lambda,\alpha}(5)$, def\/ined on the cotangent bundle
$T^{*}\mathbb{R}^5$ by~\eqref{k1}.
We will discuss various geometric structures of $T^{*}\mathbb{R}^5\setminus {\cal J}^{-1}(0)$ crucial for the
integration of Hamiltonian system def\/ined by the Hamiltonian $h:= H\circ {\cal J}: T^{*}\mathbb{R}^5\setminus {\cal
J}^{-1}(0) \rightarrow \mathbb{R}$ presented in~\eqref{e4}.

Due to the assumption $\alpha\lambda\neq 0$ we consider the matrix Lie group
\begin{gather*}
{\rm SO}_{\lambda,\alpha}(5)=\big\{g\in {\rm Mat}_{5\times 5}(\mathbb{R}):g^{\top}\eta_{\lambda,\alpha} g=\eta_{\lambda,\alpha}\big\},
\end{gather*}
where
\begin{gather*}
\eta_{\lambda, \alpha}=\left(
\begin{array}{@{}cc|c@{}} \alpha \lambda & 0 & \vec{0}^{\top}
\\
0 &\lambda & \vec{0}^{\top}
\\
\hline
\vec{0} & \vec{0} & {\boldsymbol 1}\tsep{1pt}
\end{array}
\right).
\end{gather*}

We introduce the canonical Hamiltonian action of ${\rm SO}_{\lambda,\alpha}(5)$ on the cotangent vector bundle~$
T^{*}\mathbb{R}^5$ with canonical symplectic form $\mathrm{d} \gamma$, def\/ined for $g\in {\rm SO}_{\lambda,\alpha}(5)$ and
$(q,p)\in T^{*}\mathbb{R}^5\cong \mathbb{R}^5 \times \mathbb{R}^{5^*}$~by
\begin{gather}
\label{k1}
\Phi_g\left(
\begin{matrix}
q
\\
p
\end{matrix}
\right):=\left(
\begin{array}
{@{}c|c@{}}
g {\boldsymbol 1}_5 & 0
\\
\hline
0 &(g^{-1})^{\top} {\boldsymbol 1}_5\tsep{2pt}
\end{array}
\right)\left(
\begin{matrix}
q
\\
p
\end{matrix}
\right),
\end{gather}
where ${\boldsymbol 1}_5$ is unit $5\times 5$~matrix and
\begin{gather}
\label{x1}
\gamma=p_{-1}\mathrm{d} q_{-1}+p_0\mathrm{d} q_0+\vec{p}\cdot \mathrm{d}\vec{q}.
\end{gather}
Let us note that~\eqref{k1} is the lift of the action of ${\rm SO}_{\lambda,\alpha}(5)$ from the base space~$\mathbb{R}^5$ to
the cotangent bundle $ T^{*}\mathbb{R}^5$.
So,~\eqref{k1} is a~Hamiltonian action,
see, e.g.,~\cite[Chapter~IV, Proposition~1.19]{21}.
In~\eqref{x1} we used the following notation $q^{\top}=(q_{-1},q_0,\vec{q}^{\top})$ and
$p^{\top}=(p_{-1},p_0,\vec{p}^{\top})$.
Since the case $\lambda\alpha\neq 0$ is considered, instead of the pairing~\eqref{5} we will use a~non-degenerate
pairing
\begin{gather}
\label{t1}
\mathfrak{so}(5) \times \mathfrak{\widetilde{so}}_{\lambda,\alpha}(5)\ni (\varrho, {\bf Y})\longrightarrow
\dfrac{1}{2}\Tr  (\eta_{\lambda, \alpha} {\bf Y} \varrho ) \in\mathbb{R},
\end{gather}
where $\mathfrak{so}(5)=\{\varrho\in {\rm Mat}_{5\times 5}(\mathbb{R}): \varrho^{\top}+\varrho=0\}$ and
$\mathfrak{\widetilde{so}}_{\lambda,\alpha}(5)=\{{\bf Y}\in {\rm Mat}_{5\times 5}(\mathbb{R}):  (\eta_{\lambda,
\alpha}{\bf Y} )^{\top}+ \eta_{\lambda, \alpha} {\bf Y}=0\}$ is the Lie algebra of ${\rm SO}_{\lambda,\alpha}(5)$.
Using~\eqref{t1} we will identify $\mathfrak{so}_{\lambda,\alpha}(5)^*$ with $\mathfrak{so}(5)$.
Note here that one has isomorphisms $\widetilde{{\cal \iota}}: \mathfrak{so}_{\lambda,\alpha}(5) \longrightarrow
\mathfrak{\widetilde{so}}_{\lambda,\alpha}(5)$ and ${\cal \iota}:{\cal L_+}(5) \longrightarrow \mathfrak{so}(5)$ given~by
\begin{gather}
{\bf Y}= \eta_{\lambda, \alpha}^{-1} {\bf X} \eta_{\lambda, \alpha},
\qquad
\label{s11}
\varrho= \eta_{\lambda, \alpha}^{-1} \kappa -\big(\eta_{\lambda, \alpha}^{-1} \kappa \big)^{\top}
\end{gather}
respectively, which intertwine the pairings~\eqref{5} and~\eqref{t1}.

From the identity
\begin{gather*}
\Tr\big(\eta_{\lambda, \alpha} g{\bf Y}g^{-1} \varrho\big)=\Tr \big(\eta_{\lambda, \alpha}{\bf Y}g^{-1} \varrho\big(g^{-1}\big)^{\top} \big),
\end{gather*}
we f\/ind that
\begin{gather*}
{\rm Ad}^*_{g^{-1}} \varrho = g \varrho g^{\top}
\end{gather*}
for $g\in {\rm SO}_{\lambda,\alpha}(5)$.
Now we def\/ine momentum map ${\cal{J}}: T^{*}\mathbb{R}^5\rightarrow \mathfrak{so}(5)$ as the Pl\"{u}cker map
\begin{gather}
\label{e3a1a}
{\cal{J}}(q,p):= q \big(\eta_{\lambda, \alpha}^{-1}p\big)^{\top}-\big(\eta_{\lambda, \alpha}^{-1}p\big) q^{\top},
\end{gather}
where $q, \eta_{\lambda, \alpha}^{-1} p\in \mathbb{R}^5$ and
\begin{gather*}
{\cal{J}}\circ \Phi_g={\rm Ad}^*_{g^{-1}}\circ {\cal{J}}
\end{gather*}
for $g\in {\rm SO}_{\lambda,\alpha}(5)$.
We f\/ind from~\eqref{s11} and~\eqref{e3a1a} that
\begin{gather}
a= \alpha q_{-1}p_0-q_{0}p_{-1},
\qquad
\vec{x}=\alpha\lambda q_{-1}\vec{p}-p_{-1}\vec{q},
\qquad
\vec{y}=\lambda q_{0}\vec{p}-p_{0}\vec{q},
\qquad
\vec{\mu}=\vec{q}\times \vec{p}.
\label{e3a1}
\end{gather}

A Hamiltonian action of the group ${\rm SL}(2,\mathbb{R})$ on $(T^{*}\mathbb{R}^5, \mathrm{d}\gamma)$ is def\/ined~by
\begin{gather*}
\Psi_A\left(
\begin{matrix}
q
\\
p
\end{matrix}
\right):=\left(
\begin{array}{@{}c|c@{}}
a{\boldsymbol 1}_5 & b \eta^{-1}_{\lambda, \alpha}
\\
\hline
c\eta_{\lambda, \alpha} & d {\boldsymbol 1}_5
\end{array}
\right)\left(
\begin{matrix}
q
\\
p
\end{matrix}
\right),
\end{gather*}
where $A=\left(
\begin{matrix}
a~& b
\\
c & d
\end{matrix}
\right)\in {\rm SL}(2,\mathbb{R})$.
The map $ {\cal{I}}: T^{*}\mathbb{R}^5\rightarrow \mathfrak{sl}(2,\mathbb{R}) \simeq \mathfrak{sl}(2,\mathbb{R})^* $
given by
\begin{gather}
\label{e3a3}
{\cal{I}}(q,p)= \left(
\begin{matrix}
d_3 & -d_1
\\
d_2 & -d_3
\end{matrix}
\right):= \left(
\begin{matrix}
q_{-1}p_{-1}+q_0p_0+\vec{q}\cdot\vec{p} & -\left(\alpha\lambda q_{-1}^2+\lambda q_0^2+\vec{q}\,{}^2\right)
\\
\frac{1}{\alpha\lambda}\left(p^2_{-1}+\alpha p_0^2+\alpha\lambda\vec{p}\,{}^2\right) &
-\left(q_{-1}p_{-1}+q_0p_0+\vec{q}\cdot\vec{p}\right)
\end{matrix}
\right)
\end{gather}
is an equivariant map for this action, i.e.\
\begin{gather*}
{\cal{I}}\circ \Psi_A={\rm Ad}^*_{A^{-1}}\circ {\cal{I}},
\end{gather*}
where
\begin{gather*}
{\rm Ad}^*_{A^{-1}}(\vec{d})=\left(
\begin{matrix}
a~& b
\\
c & d
\end{matrix}
\right) \left(
\begin{matrix}
d_3 & -d_1
\\
d_2 & -d_3
\end{matrix}
\right) \left(
\begin{matrix}
a~& b
\\
c & d
\end{matrix}
\right)^{-1}.
\end{gather*}
So, $ {\cal{I}}: T^{*}\mathbb{R}^5\rightarrow \mathfrak{sl}(2,\mathbb{R})$ is a~momentum map.
As usual the vector space isomorphism of Lie algebra $\mathfrak{sl}(2,\mathbb{R})$ with its dual
$\mathfrak{sl}(2,\mathbb{R})^*$ is def\/ined by the trace.
Let us recall that Lie--Poisson bracket for $\mathfrak{sl}(2,\mathbb{R})$ is given by the formula
\begin{gather}
\{f,g\}_{\mathfrak{sl}(2,\mathbb{R})}=2 d_3 \left(\frac{\partial f}{\partial d_1}\frac{\partial g}{\partial d_2}-
\frac{\partial f}{\partial d_2}\frac{\partial g}{\partial d_1}\right)+d_1 \left(\frac{\partial f}{\partial
d_1}\frac{\partial g}{\partial d_3}- \frac{\partial f}{\partial d_3}\frac{\partial g}{\partial d_1}\right)+
\nonumber
\\
\phantom{\{f,g\}_{\mathfrak{sl}(2,\mathbb{R})}=}
{}+ d_2 \left(\frac{\partial f}{\partial d_3}\frac{\partial g}{\partial d_2}- \frac{\partial f}{\partial d_2}\frac{\partial g}{\partial d_3}\right).
\label{5a1}
\end{gather}

\begin{Proposition}
For both momentum maps mentioned above the following holds:
\begin{enumerate}[$(i)$]\itemsep=0pt
\item They prove to be Poisson maps, i.e.~arrows in the diagram
\begin{gather}
\label{d1}
\begin{split}&
\xymatrix{
 & & T^{*}\mathbb{R}^5 \ar@<.0ex>[ddrr]^*-<.9ex>\txt{{{$\mathcal{J}$}}}
             \ar@<-.0ex>[ddll]^*-<4ex>\txt{{{$\mathcal{I}$}}}      & &
             &  \\
             & &&&    &\\
  \mathfrak{sl}(2,\mathbb{R})    &&&     &   \mathfrak{so}(5)
 }
\end{split}
\end{gather}
are morphisms of Poisson manifolds.
\item The momentum maps' fibers ${\cal{I}}^{-1}(\vec{d})$ and ${\cal{J}}^{-1}(\rho)$ over
$\vec{d}=(d_1,d_2,d_3)^{\top}\in \mathfrak{sl}(2,\mathbb{R})$ and $\rho\in \mathfrak{so}(5)$ are symplectically
orthogonal, i.e.\
\begin{gather}
\label{s7}
\{{{\cal{I}}^{*}(C^{\infty}({\mathfrak{sl}(2,\mathbb{R}}))), \cal{J}}^{*}(C^{\infty}(\mathfrak{so}(5))) \}=0,
\end{gather}
where $\{\cdot, \cdot \}$ is the canonical Poisson bracket on $T^{*}\mathbb{R}^5$.
\end{enumerate}
\end{Proposition}
\begin{proof}
The property~\eqref{s7} follow from Leibniz rule and relations
\begin{gather*}
\{d_k,a\}=\{d_k,\vec{\mu}\}=\{d_k,\vec{x}\}=\{d_k,\vec{y}\}=0,
\end{gather*}
where $d_k$ and $a$, $\vec{\mu}$, $\vec{x}$, $\vec{y}$ are given by~\eqref{e3a3} and~\eqref{e3a1}, respectively.
\end{proof}

From the above properties of ${\cal{I}}$ and ${\cal{J}}$ we conclude that diagram~\eqref{d1} realizes symplectic dual
pair.
For the def\/inition of symplectic dual pair see~\cite[Chapter~IV, Section~9.3]{11}.

We will consider $ T^{*}\mathbb{R}^5$ as union of two complementary subsets
\begin{gather*}
T^{*}\mathbb{R}^5= T_{\sing}^{*}\mathbb{R}^5\cup T_{\reg}^{*}\mathbb{R}^5,
\end{gather*}
where the subset $ T_{\sing}^{*}\mathbb{R}^5$ consists of the pairs $(q,p)\in T_{\sing}^{*}\mathbb{R}^5$ such that
$q\in\mathbb{R}^5$ and $\eta_{\lambda, \alpha}^{-1} p\in \mathbb{R}^5$ are linearly dependent
while $(q,p)\in T_{\reg}^{*}\mathbb{R}^5$ if\/f~$q$ and $\eta_{\lambda, \alpha}p$ are linearly independent.
Note that $T_{\sing}^{*}\mathbb{R}^5={\cal J}^{-1}(0)$ and so, it is closed in $ T^{*}\mathbb{R}^5$.

The function
\begin{gather}
\label{v1}
c:= \det {\cal I}(q,p)= d_1d_2- d_3^2
\end{gather}
is a~Casimir of the Poisson bracket~\eqref{5a1} and the equality
\begin{gather}
\label{v3}
\delta_{\lambda,\alpha}:= c\circ{\cal I}=\frac{1}{\alpha \lambda} \big(c_1\circ \iota^{-1} \circ {\cal J}\big),
\end{gather}
is valid, where $c_1$ is Casimir function def\/ined in~\eqref{ab55}.
See~\eqref{s11} for def\/inition of ${\cal \iota}:{\cal L_+}(5) \longrightarrow \mathfrak{so}(5)$.
The function $\delta_{\lambda,\alpha}$ as well as the subsets $T_{\sing}^{*}\mathbb{R}^5$ and $T_{\reg}^{*}\mathbb{R}^5$
are invariant with respect to the action of the groups $\Phi  ({\rm SO}_{\lambda,\alpha}(5) )$ and
$\Psi ({\rm SL}(2,\mathbb{R}) )$.
Let us also mention that
\begin{gather*}
\Psi_A\circ \Phi_g=\Phi_g\circ \Psi_A
\end{gather*}
for $A\in {\rm SL}(2,\mathbb{R})$ and $g\in {\rm SO}_{\lambda,\alpha}(5)$.

We will present other important facts in the following
\begin{Proposition}\label{proposition3.2}\quad
\begin{enumerate}[$(i)$]\itemsep=0pt
\item For $A\in {\rm GL}(2,\mathbb{R})$ one has
\begin{gather}
\label{s2}
\left({\cal{I}}\circ \Psi_A\right)(q,p) {\boldsymbol \epsilon} =A {\cal{I}}(q,p) {\boldsymbol \epsilon} A^{\top},
\end{gather}
where $(q,p)\in \mathbb{R}\times \mathbb{R}^{5^*}$ and ${\boldsymbol \epsilon}=\left(
\begin{matrix}
0 & 1
\\
-1 & 0
\end{matrix}
\right)$.
\item The fibres $\Gamma_s:=\delta_{\lambda,\alpha}^{-1}\left(s\right)$, $s\in\mathbb{R}$, of $\delta_{\lambda,\alpha}:
T^*\mathbb{R}^5\rightarrow \mathbb{R}$ are $9$-dimensional submanifolds of $T^*_{\reg}\mathbb{R}^5$ invariant with
respect to the subgroup ${\rm SL}_{\pm}(2,\mathbb{R})\subset {\rm GL}(2,\mathbb{R})$, consisting of such $A\in {\rm GL}(2,\mathbb{R})$
that $\det A=\pm 1$; they are also invariant with respect to the group ${\rm SO}_{\lambda,\alpha}(5)$.
\item The fibres ${\cal I}^{-1}(\vec{d})$ of ${\cal{I}}:T^*\mathbb{R}^5\rightarrow \mathfrak{sl}(2,\mathbb{R})$,
$\vec{d}\in \mathbb{R}^3 \cong \mathfrak{sl}(2,\mathbb{R})$ defined by equations
\begin{gather}
\label{ee11}
  d_1=\alpha\lambda q_{-1}^2+\lambda q_0^2+\vec{q}\,{}^2,
\\
\label{ee12}
  d_2=\dfrac{1}{\alpha\lambda}p^2_{-1}+\dfrac{1}{\lambda} p_0^2+\vec{p}\,{}^2,
\\
\label{ee13}
  d_3=q_{-1}p_{-1}+q_0p_0+(\vec{q}\cdot\vec{p})
\end{gather}
are $7$-dimensional submanifolds of $T^*_{\reg}\mathbb{R}^5$.
They are also invariant with respect to the action of ${\rm SO}_{\lambda, \alpha}(5)$ and the action of stabilizer subgroup
${\rm SL}(2,\mathbb{R})_{\vec{d}}$.
\end{enumerate}
\end{Proposition}

\begin{proof}
Equivariance property~\eqref{s2} and the facts that f\/ibres $\Gamma_s=\delta_{\lambda,\alpha}^{-1} (s )$ and
${\cal I}^{-1}(\vec{d})$, for $\vec{d}\neq \vec{0}$, are submanifolds of $T^*_{\reg}\mathbb{R}^5$ can be easily verif\/ied
by the direct calculations.
From~\eqref{v1} and~\eqref{s2} one obtains
\begin{gather*}
 (\delta_{\lambda,\alpha}\circ \Psi_A )(q,p) = (\det A )^2 \delta_{\lambda,\alpha}(q,p).
\end{gather*}
So, submanifold $\Gamma_s\subset T^*_{\reg}\mathbb{R}^5$ is invariant with respect to $\Psi ({\rm SL}_{\pm}(2,\mathbb{R}) )$.
\end{proof}

For $A\in {\rm GL}(2,\mathbb{R})$ one has
\begin{gather*}
{\cal{J}}  (\Psi_A(q,p) ) =\det A
{\cal{J}}(q,p).
\end{gather*}
Thus according to the theory of Grassmannians,
see, e.g.,~\cite[Chapter~I, Section~5]{25}, we note that the momentum map~\eqref{e3a1a} def\/ines the Pl\"{u}cker embeding
${\cal{P}}:G(2,5)\rightarrow \mathbb{P}(\bigwedge^2\mathbb{R}^5)\cong \mathbb{P}(\mathfrak{so}(5))$ of the
Grassmannian $G(2,5)$ of the $2$-dimensional vector subspaces of $\mathbb{R}^5$, spanned by vectors
$q,\eta^{-1}_{\lambda,\alpha}p\in\mathbb{R}^5$.
Thus the image ${\cal J}(T_{\reg}^{*}\mathbb{R}^5)$ of $T_{\reg}^{*}\mathbb{R}^5$ in $\mathfrak{so}(5)$ is described by the Pl\"{u}cker relations
\begin{gather}
\lambda a\vec{\mu}-\vec{x}\times \vec{y}=0,
\qquad
\vec{\mu}\cdot \vec{x}=0,
\qquad
\label{x2x}
\vec{\mu}\cdot \vec{y}=0,
\end{gather}
which one obtains directly from~\eqref{e3a1}.
We also observe that $T^*_{\reg}\mathbb{R}^5$ has structure of the ${\rm GL}(2,\mathbb{R})$-principal bundle, i.e.~it is the
total space of Stiefel principal bundle
\begin{gather}
\label{d1111x}
\begin{split}  &
\xymatrix{
  {\rm GL}(2,\mathbb{R}) \ar@<-.0ex>[r]^*+<1ex>\txt{{{\cal{}}}}  &  T_{\reg}^{*}\mathbb{R}^5   \ar@<.0ex>[dd]^*+<1ex>\txt{{{$\mathcal{\pi}$}}}&&\\
    &&    &\\
 &  G(2,5) &&\\
}
\end{split}
\end{gather}
over $ G(2,5)$, for def\/inition of Stiefel bundle see~\cite{5axc}.
Equations~\eqref{x2x}
def\/ine $7$-dimensional submanifold ${\cal J}\left(T_{\reg}^{*}\mathbb{R}^5 \right)$ in
$\mathfrak{so}(5)$ which is invariant with respect to the multiplication $\mathfrak{so}(5) \ni \varrho\rightarrow
r\varrho\in \mathfrak{so}(5)$ of~$\varrho$ by $r\in \mathbb{R}\setminus \{0\}$.
So, one has the $\left(\mathbb{R}\setminus \{0\}\right)$-principal bundle
\begin{gather}
\label{d1111xx}
\begin{split}
& \xymatrix{
 \mathbb{R}\setminus\{0\}\ar@<-.0ex>[r]^*+<1ex>\txt{{{\cal{}}}} & {\cal J}
\left(T_{\reg}^{*}\mathbb{R}^5\right)\ar@<.0ex>[dd]^*+<1ex>\txt{{{$\mathcal{}$}}}&&\\
   &&    &\\
 & {\cal P}\left( G(2,5)\right)\cong G(2,5) &&\\
}
\end{split}
\end{gather}
over the Grassmannian $G(2,5)$.
Let us note here that~\eqref{d1111xx} is the determinant bundle of the bundle~\eqref{d1111x}.
Thus one has the surjective morphism
\begin{gather*}
\xymatrix{
  T_{\reg}^{*}\mathbb{R}^5 \ar@<.0ex>[dd]^*+<1ex>\txt{ {{$\mathcal{\pi}$}}}
\ar@<-.0ex>[r]^*+<1ex>\txt{ {{$\mathcal{J}$}}}  &  {\cal J}\left(T_{\reg}^{*}\mathbb{R}^5\right)
\ar@<.0ex>[dd]^*+<1ex>\txt{ {{$\mathcal{\widetilde{\pi}}$}}}&&\\
 &&    &\\
  G(2,5)  \ar@<-.0ex>[r]^*+<1ex>\txt{{{$\mathcal{}$}}}  &  G(2,5) &&\\
}
\end{gather*}
of the principal bundles def\/ined by the momentum map ${\cal J}: T_{\reg}^{*}\mathbb{R}^5 \rightarrow \mathfrak{so}(5)$
and the determinant map $\det: {\rm GL}(2,\mathbb{R})\rightarrow \mathbb{R}\setminus \{0\} $.

Since, submanifold $\Gamma_s\subset T^*_{\reg}\mathbb{R}^5$ is invariant with respect to the action of
${\rm SL}_{\pm}(2,\mathbb{R})$ it is a~total space of the ${\rm SL}_{\pm}(2,\mathbb{R}) $-principal subbundle of the
${\rm GL}(2,\mathbb{R}) $-principal bundle~\eqref{d1111x}.
The structural groups morphism in this case is given by the inclusion ${\rm SL}_{\pm}(2,\mathbb{R})\hookrightarrow
{\rm GL}(2,\mathbb{R}) $.

On the other hand submanifold $\Omega_s:= {\cal J}(T_{\reg}^{*}\mathbb{R}^5)\cap c_1^{-1}(\alpha\lambda
s)\subset \mathfrak{so}(5)$ is total space of a~$\mathbb{Z}_2$-principal bundle over $G(2,5)$.
In the subsequent diagram we present the morphisms of the principal bundles mentioned above
\begin{gather*}
\xymatrix{
  \Omega_s  \ar@<.0ex>[dd]^*+<1ex>\txt{{{$\mathcal{\widetilde{\pi}}_s$}}}
 & \Gamma_s \ar@<.0ex>[dd]^*+<1ex>\txt{{{$\mathcal{\pi}_s$}}} \ar@<-.0ex>[r]^*+<1ex>\txt{{{$l$}}}
\ar@<-.0ex>[l]_*+<1ex>\txt{{{$  \mathcal{J}$}}}
 &    T_{\reg}^{*}\mathbb{R}^5  \ar@<.0ex>[dd]^*+<1ex>\txt{{{$\mathcal{\pi}$}}}&&\\
     &&&    &\\
  G(2,5) \ar@<-.0ex>[r]^*+<1ex>\txt{{{$\rm id$}}} & G(2,5)  \ar@<-.0ex>[r]^*+<1ex>\txt{{{$\rm id$}}}  &  G(2,5) &&\\
}
\end{gather*}
The corresponding structural group epimorphism for ${\cal J}: \Gamma_s \rightarrow \Omega_s$ is $\det:
{\rm SL}_{\pm}(2,\mathbb{R}) \rightarrow \mathbb{Z}_2=\{-1,1\}$.
The bundle map ${\cal J}: \Gamma_s \rightarrow \Omega_s$ is a~surjective submersion and bundle projection
$\widetilde{\pi}_s: \Omega_s \rightarrow G(2,5)$ def\/ines a~two-fold covering of the Grassmannian $G(2,5)$.

{\sloppy One has the decompositions ${\rm GL}(2,\mathbb{R})={\rm GL}_{2}(2,\mathbb{R}) \cdot {\rm GL}_{+}(2,\mathbb{R})$ and
${\rm SL}_{\pm}(2,\mathbb{R})={\rm GL}_2(2,\mathbb{R}) \cdot {\rm SL}(2,\mathbb{R})$, where ${\rm GL}_{2}(2,\mathbb{R}):=\Big\{{\mbox{\scriptsize $\left(
\begin{matrix}
1 & 0
\\
0 & 1
\end{matrix}
\right)$}}, {\mbox{\scriptsize$\left(
\begin{matrix}
0 & 1
\\
1 & 0
\end{matrix}
\right)$}} \Big\} \cong \mathbb{Z}_2$ and ${\rm GL}_{+}(2,\mathbb{R}):= \{A\in {\rm GL}(2,\mathbb{R})$: \mbox{$\det A>0 \}$}.
The map $\psi_{\mbox{\scriptsize$\left(
\begin{matrix}
0 & 1
\\
1 & 0
\end{matrix}
\right)$}}:T^*_{\reg}\mathbb{R}^5 \rightarrow T^*_{\reg}\mathbb{R}^5$ changes the orientation of the frame def\/ined by the
pair of vectors $(q,\eta_{\lambda,\alpha}^{-1} p)$ and $\bigg({\cal J}\circ \psi_{\mbox{\scriptsize$\left(
\begin{matrix}
0 & 1
\\
1 & 0
\end{matrix}
\right)$}}\bigg)(q,p) = - {\cal J} (q,p)$.
Hence we can consider $\Omega_s\cong G_{+}(2,5)$ as the Grassmannian of $2$-dimensional subspaces in $\mathbb{R}^5$
with f\/ixed orientation.

}

One has the double principal bundle structure on $\Gamma_s$ with structural groups ${\rm SL}(2,\mathbb{R})$ and ${\rm SO}_{\lambda,\alpha}(5)$ and momentum maps ${\cal I}$ and ${\cal J}$ being bundle projections
\begin{gather}
\label{d111s5}
\begin{split}&
\xymatrix{
 & & \Gamma_s \ar@<.0ex>[ddrr]^*-<.4ex>\txt{{{$\mathcal{J}$}}}
             \ar@<-.0ex>[ddll]^*-<4.4ex>\txt{{{$\mathcal{I}$}}}      & &
             &  \\
             & &&&    &\\
 \Delta_s   &&&    &   \Omega_s
 }
\end{split}
\end{gather}
where $\Delta_s:=c^{-1}(s)$ and $s\in\mathbb{R}$.
For $s=0$ we assume by def\/inition that $\vec{0}\notin \Delta_0$.
The restriction $\mathrm{d} \gamma\big|_{\Gamma_s}$ of symplectic form $\mathrm{d} \gamma$ to $\Gamma_s$ is invariant
with respect to $\Psi ({\rm SL}_{\pm}(2,\mathbb{R})  )$ and $\Phi  ({\rm SO}_{\lambda,\alpha}(5) )$.
So, applying reduction procedure to both these actions one obtains
the reduced symplectic manifolds $\Gamma_s/{\rm SO}_{\lambda,\alpha}(5) \cong \Delta_s$ and $\Gamma_s/{\rm SL}(2,\mathbb{R}) \cong \Omega_s$.

The Hamiltonian f\/low $\{\sigma_t^{\lambda,\alpha}\}_{t\in\mathbb{R}}$ on $ T_{\reg}^{*}\mathbb{R}^5$ def\/ined
by the Hamiltonian $\delta_{\lambda,\alpha}=\frac{1}{\alpha\lambda}(c_1\circ \iota^{-1} \circ {\cal J})$ is
described explicitly by expressions~\eqref{n1},~\eqref{s6} established in Section~\ref{Section4}.
It preserves f\/ibres ${\cal J}^{-1}(\varrho)$ and ${\cal I}^{-1}(\vec{d})$ of both momentum maps
and on ${\cal I}^{-1}(\vec{d})$ it is identical to the action
of the stabilizer subgroup ${\rm SL}(2,\mathbb{R})_{\vec{d}}\subset {\rm SL}(2,\mathbb{R})$.
From~\eqref{v3} one sees that $\delta_{\lambda,\alpha}$ is the pull-back of the Casimirs~$c$ and
$\frac{1}{\alpha\lambda} (c_1\circ \iota^{-1}  )$.
Thus the groups $ {\rm SL}(2,\mathbb{R})$ and ${\rm SO}_{\lambda,\alpha}(5)$ act also on the reduced symplectic manifold
$\widetilde{\Gamma}_s:= \Gamma_s/\{\sigma_t^{\lambda,\alpha}\}$ by symplectomorphisms.
Summarizing the above facts we can formulate
\begin{Proposition}
For any $s\in \mathbb{R}$ one has the symplectic double fibration
\begin{gather}
\label{d111s6}
\begin{split}&
\xymatrix{
 & & \widetilde{\Gamma}_s \ar@<.0ex>[ddrr]^*-<.4ex>\txt{{{$\mathcal{ \widetilde{J}}$}}}
             \ar@<-.0ex>[ddll]^*-<4.8ex>\txt{{{$\mathcal{ \widetilde{I}}$}}}      & &
             &  \\
             & &&&    &\\
 \Delta_s   &&&    &   \Omega_s
}
\end{split}
\end{gather}
i.e.~all manifolds in~\eqref{d111s6} are symplectic and the maps $\mathcal{\widetilde{I}}$
and $\mathcal{\widetilde{J}}$ are surjective Poisson submersions.
Moreover $\mathcal{\widetilde{I}}$-fibres are symplectically orthogonal to the $\mathcal{\widetilde{J}}$-fibres.
\end{Proposition}

\begin{proof}
The symplectic orthogonality of $\mathcal{\widetilde{I}}$-f\/ibres and $\mathcal{\widetilde{J}}$-f\/ibres follows
from~\eqref{s7}.
Due to the fact that both momentum maps are constant on the trajectories of
$\{\sigma_t^{\lambda,\alpha}\}_{t\in\mathbb{R}}$ the surjective epimorphisms $\mathcal{\widetilde{I}}$ and
$\mathcal{\widetilde{J}}$ are def\/ined by $\mathcal{I}$ and $\mathcal{J}$ respectively.
\end{proof}

\begin{Remark}
In general the f\/ibres of $\mathcal{\widetilde{I}}$ and $\mathcal{\widetilde{J}}$ are neither connected nor simply
connected.
So, symplectic manifolds $\Delta_s$ and $\Omega_s$ are not Morita equivalent in sense of~\cite[Chapter~IV, Section~9.3]{11}.
\end{Remark}

Since $\Omega_s\subset \mathfrak{so}(5)$ is invariant with respect to the coadjoint action of ${\rm SO}_{\lambda,\alpha}(5)$
we will investigate the decomposition of $\Omega_s$ into the orbits of this action.
For this reason we note that $ ({\boldsymbol \epsilon} {\cal I} )^{-1}(\vec{d})\subset T^*_{\reg}\mathbb{R}^5$,
where $ ({\boldsymbol \epsilon} {\cal I} )(q,p):={\boldsymbol \epsilon} {\cal I}(q,p)$, is invariant with
respect to the action~\eqref{k1}.
From  Proposition~\ref{proposition3.2}(i)
one has
\begin{gather}
\label{p1aaacc}
A\left(
\begin{matrix}
d_1(q,p) & d_3(q,p)
\\
d_3(q,p) & d_2(q,p)
\end{matrix}
\right) A^{\top}= \left(
\begin{matrix}
d_1\left(\Psi_A (q,p)\right) & d_3\left(\Psi_A (q,p)\right)
\\
d_3\left(\Psi_A (q,p)\right) & d_2\left(\Psi_A (q,p)\right)
\end{matrix}
\right)
\end{gather}
for $A\in {\rm SL}(2,\mathbb{R})$.
Since both maps in diagram~\eqref{d111s5} are surjective submersion we f\/ind that $\Omega_s={\cal J}(\Gamma_s
)= {\cal J}(({\boldsymbol \epsilon} {\cal I})^{-1}(\Delta_s) )$.
It follows from~\eqref{p1aaacc} that ${\cal J}(({\boldsymbol \epsilon} {\cal
I})^{-1}(\vec{d}) ) = {\cal J}(({\boldsymbol \epsilon} {\cal
I})^{-1}(\vec{d}') )$ if\/f for $\vec{d},\vec{d}'\in\Delta_s\subset {\rm SL}(2,\mathbb{R})$ there exists
$A\in {\rm SL}(2,\mathbb{R})$ such
\begin{gather*}
\left(
\begin{matrix}
d'_1 & d'_3
\\
d'_3 & d'_2
\end{matrix}
\right)= A\left(
\begin{matrix}
d_1 & d_3
\\
d_3 & d_2
\end{matrix}
\right) A^{\top}.
\end{gather*}
So, in order to describe invariant subsets ${\cal J}(({\boldsymbol \epsilon} {\cal
I})^{-1}(\vec{d}) \cap T^*_{\reg}\mathbb{R}^5 )\subset \Omega_s$, where $\vec{d}\in\Delta_s$, we
formulate
\begin{Proposition}\label{Proposition3.5}
For any $\vec{d}\in\Delta_s$ there is $A\in {\rm SL}(2,\mathbb{R})$ such that:
\begin{enumerate}[$(i)$]\itemsep=0pt
\item if $s<0$
\begin{gather*}
\left(
\begin{matrix}
d_1 & d_3
\\
d_3 & d_2
\end{matrix}
\right)=A\left(
\begin{matrix}
s & 0
\\
0 & 1
\end{matrix}
\right) A^{\top},
\end{gather*}
\item if $s\geq 0$
\begin{gather*}
\left(
\begin{matrix}
d_1 & d_3
\\
d_3 & d_2
\end{matrix}
\right)=\pm A\left(
\begin{matrix}
s & 0
\\
0 & 1
\end{matrix}
\right) A^{\top}.
\end{gather*}
\end{enumerate}
If $(q,p)\in  ({\boldsymbol \epsilon} {\cal I} )^{-1} (\vec{d} ) \cap T^*_{\reg}\mathbb{R}^5 $ then the
signature of the symmetric form $\left(
\begin{matrix}
d_1 & d_3
\\
d_3 & d_2
\end{matrix}
\right)$ is the same as the signature of the restriction $\eta_{\lambda,\alpha}\big|_{V}$ of $\eta_{\lambda,\alpha}$ to
the $2$-dimensional subspace $V\subset \mathbb{R}^5$ spaned by~$q$ and $\eta_{\lambda,\alpha}^{-1} p$.
The action~\eqref{k1} preserves the signature of $\eta_{\lambda,\alpha}\big|_{V}$.
\end{Proposition}

Let us note that if $V_1,V_2\in G_+(2,5)$ have identical signatures with respect to $\eta_{\lambda,\alpha}$ then they
belong to the same orbit of ${\rm SO}_{\lambda,\alpha}(5)$.
Thus and from Proposition~\ref{Proposition3.5} we conclude that the following proposition is valid.
\begin{Proposition}\label{Proposition3.6}\quad
\begin{enumerate}[$(i)$]\itemsep=0pt
\item If $s<0$ then $\Omega_s$ is six-dimensional ${\rm Ad}^* ({\rm SO}_{\lambda,\alpha}(5) )$-orbit which is isomorphic
$($as a~homogeneous space$)$ to the Grassmannian $G^{+-}_{+}(2,5)$ of the $2$-dimensional
oriented subspaces $V\subset\mathbb{R}^5$ such that $\sign\eta_{\lambda,\alpha}\big|_{V}=(+-)$.
\item If $s=0$ then $\Omega_0$ is decomposed into the six-dimensional ${\rm Ad}^* ({\rm SO}_{\lambda,\alpha}(5) )$-orbits
$\Omega_0^{+0}$ and $\Omega_0^{-0}$ which are isomorphic to the Grassmannians $G^{+0}_{+}(2,5)$ and $G^{-0}_{+}(2,5)$ of
the $2$-dimensional oriented subspaces $V\subset \mathbb{R}^5$ such that $\sign\eta_{\lambda,\alpha}\big|_{V}$ are $(+0)$ and $(-0)$, respectively.
\item If $s>0$ then $\Omega_s$ is decomposed into the six-dimensional ${\rm Ad}^* ({\rm SO}_{\lambda,\alpha}(5) )$-orbits
$\Omega_s^{++}$ and $\Omega_s^{--}$ which are isomorphic to the Grassmannians $G^{++}_{+}(2,5)$ and $G^{--}_{+}(2,5)$ of
the $2$-dimen\-sional oriented subspaces $V\subset \mathbb{R}^5$ such that $\sign\eta_{\lambda,\alpha}\big|_{V}$
are $(++)$ and $(--)$, respectively.
\item If $\vec{d}=\vec{0}$ then ${\cal J}(({\boldsymbol \epsilon} {\cal I})^{-1}(\vec{0}))$
is a~four-dimensional ${\rm Ad}^*({\rm SO}_{\lambda,\alpha}(5))$-orbit isomorphic to the Grassmannian
$G^{00}_{+}(2,5)$ of the $2$-dimensional oriented subspaces $V\subset \mathbb{R}^5$ such that $\sign\eta_{\lambda,\alpha}\big|_{V}=(00)$.
\end{enumerate}
\end{Proposition}

In the case $\alpha\lambda\neq 0$ the group ${\rm SO}_{\lambda,\alpha}(5)$ is isomorphic to one of the following groups:
${\rm SO}(5)$, ${\rm SO}(1,4)$ and ${\rm SO}(2,4)$.
Let us describe all these subcases separately.
\begin{Proposition}\label{Proposition3.7}\quad
\begin{enumerate}[$(i)$]\itemsep=0pt
\item For special orthogonal group ${\rm SO}(5)$ one has $s>0$ and $\Omega_s\cong G^{++}_{+}(2,5)$.
\item For de Sitter group ${\rm SO}(1,4)$ one has: $\Omega_s\cong G^{++}_{+}(2,5)$ for $s>0$, $\Omega_s\cong G^{+-}_{+}(2,5)$
for $s<0$; $\Omega_0\cong G^{+0}_{+}(2,5)$ for $s=0$.
\item For anti-de Sitter group ${\rm SO}(2,3)$ one has: $\Omega_s\cong G^{++}_{+}(2,5)$ or $\Omega_s\cong G^{--}_{+}(2,5)$ for
$s>0$; $\Omega_s\cong G^{+-}_{+}(2,5)$ for $s<0$ and $\Omega_0\cong G^{+0}_{+}(2,5)\cup G^{-0}_{+}(2,5)$ for $s=0$.
The case ${\cal J}(({\boldsymbol \epsilon} {\cal I})^{-1}(\vec{0}))\cong
G^{00}_{+}(2,5)$ described in~Proposition~{\rm \ref{Proposition3.6}}$(iv)$ is admissible for the anti-de Sitter group.
\end{enumerate}
\end{Proposition}

Completing this section let us shortly discuss the case $T^*_{\sing}\mathbb{R}^5= {\cal J}^{-1}(0)$.
If $(q,p)\in {\cal J}^{-1}(0)$ then one has $b_1 q+b_2 \eta_{\lambda,\alpha}^{-1} p=0$ for some $0\neq \left(
\begin{matrix}
b_1
\\
b_2
\end{matrix}
\right)\in\mathbb{R}^2$.
Thus, we f\/ind that ${\cal I}(q,p) {\boldsymbol \epsilon} \left(
\begin{matrix}
b_1
\\
b_2
\end{matrix}
\right) =0$ and hence ${\cal I}\left({\cal J}^{-1}(0)\right)\subset \Delta_0$.
Summing up the above facts we obtain bundle ${\cal I}: {\cal J}^{-1}(0)\setminus \{0\} \rightarrow \Delta_0 \setminus\{0\}$.

The canonical form~$\gamma$ after restriction to ${\cal J}^{-1}(0)\setminus \{0\}$ is given by
\begin{gather*}
\gamma\big|_{{\cal J}^{-1}(0)}=\frac 12d_3 \mathrm{d} \ln |d_1|=-\frac 12 d_3 \mathrm{d} \ln |d_2|+ \mathrm{d}d_3.
\end{gather*}
So, $\mathrm{d}\gamma\big|_{{\cal J}^{-1}(0)}$ is equal to the lifting ${\cal I}^*\omega_0$
of the ${\rm SL}(2,\mathbb{R})$-invariant symplectic form $\omega_0$ of the symplectic leaf $\Delta_0\subset {\rm SL}(2,\mathbb{R})$.
We will not consider this case in what follows.
The reason is that the Hamiltonian $H\circ {\cal J}$ after restriction to ${\cal J}^{-1}(0)$ vanishes, so it generates trivial dynamics.

In the next section we will use f\/ibration~\eqref{d111s5} to integrate Hamiltonian equations def\/ined by Hamiltonian
$H\circ {\cal J}$ for regular case ${\cal I}^{-1}(\vec{d})\cap T_r^{*}\mathbb{R}^5$.

\section{Solutions and their physical interpretations}\label{Section4}

Our goal is to use results of two previous section for solving Hamilton equations
\begin{gather}
\label{r1}
\frac{dq}{dt}=\frac{\partial h}{\partial p}
\qquad
\text{and}
\qquad
\frac{dp}{dt}=-\frac{\partial h}{\partial q}
\end{gather}
on $T^{*}\mathbb{R}^5$ with Hamiltonian
\begin{gather}
h:= H\circ {\cal J}=\gamma \big(\alpha \big(\alpha\lambda^2 q_{-1}^2+\lambda^2q_0^2+\epsilon\vec{q}\,{}^2\big)
\vec{p}\,{}^2+ \big(p_{-1}^2+\alpha p_0^2\big) \big(\vec{q}\,{}^2+\epsilon q_0^2+\alpha \epsilon q_{-1}^2\big)
\nonumber
\\
\phantom{h:=}{}
-\alpha\epsilon \big(q_{-1}p_{-1}+q_0p_0+\vec{q}\cdot\vec{p} \big)^2- 2\alpha(\lambda -\epsilon)
 (q_{-1}p_{-1}+q_0p_0 ) (\vec{q}\cdot\vec{p}  ) \big)
\nonumber
\\
\phantom{h:=}{}
+\nu (\lambda-\epsilon)^2  (\alpha q_{-1}p_0-q_0p_{-1} )^2  (\vec{q}\times \vec{p} )^2,
\label{e4}
\end{gather}
where~$H$ is def\/ined in~\eqref{a51}.
After substituting~\eqref{e4} into~\eqref{r1} we obtain{\samepage
\begin{gather}
\frac{dq_{-1}}{dt}=2\gamma \big(\big(\vec{q}\,{}^2+\epsilon q_0^2\big) p_{-1}- \alpha  (\epsilon q_0p_0+\lambda
\vec{q}\cdot\vec{p}  ) q_{-1}\big)
\nonumber
\\[-0.4ex]
\phantom{\frac{dq_{-1}}{dt}=}
{}-2 \nu (\lambda-\epsilon)^2  (\alpha q_{-1}p_0-q_0p_{-1} )  (\vec{q}\times \vec{p} )^2 q_0,
\nonumber
\\[-0.4ex]
\frac{dq_0}{dt}=2\gamma\alpha \big(\big(\vec{q}\,{}^2+\alpha\epsilon q_{-1}^2\big) p_{0}-  (\epsilon
q_{-1}p_{-1}+\lambda \vec{q}\cdot\vec{p}  ) q_{0}\big)
\nonumber
\\[-0.4ex]
\phantom{\frac{dq_0}{dt}=}
{}+2 \nu\alpha (\lambda-\epsilon)^2  (\alpha q_{-1}p_0-q_0p_{-1} )  (\vec{q}\times \vec{p} )^2 q_{-1},
\nonumber
\\[-0.4ex]
\frac{dp_{-1}}{dt}=-2\gamma\alpha \big(\big(\alpha\lambda^2 \vec{p}\,{}^2+\alpha\epsilon p_0^2\big) q_{-1}-
 (\epsilon q_0p_0+\lambda \vec{q}\cdot\vec{p}  ) p_{-1}\big)
\nonumber
\\[-0.4ex]
\phantom{\frac{dp_{-1}}{dt}=}
{}- 2\alpha \nu (\lambda-\epsilon)^2  (\alpha q_{-1}p_0-q_0p_{-1} )  (\vec{q}\times \vec{p} )^2 p_0,
\nonumber
\\[-0.4ex]
\frac{dp_0}{dt}=-2\gamma \big(\big(\alpha\lambda^2 \vec{p}\,{}^2+\epsilon p_{-1}^2\big) q_{0}-\alpha  (\epsilon
q_{-1}p_{-1}+\lambda \vec{q}\cdot\vec{p}  ) p_{0}\big)
\nonumber
\\[-0.4ex]
\phantom{\frac{dp_0}{dt}=}
{}+ 2 \nu (\lambda-\epsilon)^2  (\alpha q_{-1}p_0-q_0p_{-1} )  (\vec{q}\times \vec{p} )^2 p_{-1},
\nonumber
\\[-0.4ex]
\frac{d\vec{q}}{dt}=2\gamma\alpha \big(\big(\epsilon \vec{q}\,{}^2+\lambda^2 q_0^2+\alpha\lambda^2 q_{-1}^2\big)
\vec{p}-  (\lambda q_{-1}p_{-1}+\lambda q_0p_0+\epsilon \vec{q}\cdot\vec{p}  ) \vec{q}\big)
\nonumber
\\[-0.4ex]
\phantom{\frac{d\vec{q}}{dt}=}
{}+ 2 \nu (\lambda-\epsilon)^2  (\alpha q_{-1}p_0-q_0p_{-1} )^2  \big(\vec{q}\,{}^2\vec{p}- (\vec{q}\cdot
\vec{p} )\vec{q}\big),
\nonumber
\\[-0.4ex]
\frac{d\vec{p}}{dt}=-2\gamma \big(\big(\alpha\epsilon \vec{p}\,{}^2+\alpha p_0^2+ p_{-1}^2\big) \vec{q}-\alpha
 (\lambda q_{-1}p_{-1}+\lambda q_0p_0+\epsilon \vec{q}\cdot\vec{p}  ) \vec{p}\big)
\nonumber
\\[-0.4ex]
\phantom{\frac{d\vec{p}}{dt}=}
{}- 2 \nu (\lambda-\epsilon)^2  (\alpha q_{-1}p_0-q_0p_{-1} )^2 \big(\vec{p}\,{}^2\vec{q}- (\vec{q}\cdot
\vec{p} )\vec{p}\big).
\label{e7}
\end{gather}}

\noindent
Using~\eqref{e3a1} and~\eqref{ee11}--\eqref{ee13} we transform above system of equations to the following one
\begin{gather}
\label{nn1}
\frac{d}{dt}\!\left(\!
\begin{matrix}
\sqrt{\alpha} q_{-1}(t)
\\
q_0(t)
\\
p_{-1}(t)
\\
\sqrt{\alpha} p_0(t)
\end{matrix}
\!\right)=\left(\!
\begin{matrix}
-2\gamma \alpha\lambda d_3 & \sqrt{\alpha} B & 2\gamma\sqrt{\alpha} d_1 & 0
\\
-\sqrt{\alpha} B & -2\gamma \alpha\lambda d_3 & 0 & 2\gamma\sqrt{\alpha} d_1
\\
-2\gamma \alpha\sqrt{\alpha}\lambda^2 d_2\! & 0 & 2\gamma \alpha\lambda d_3 & \sqrt{\alpha} B
\\
0 & -2\gamma\alpha \sqrt{\alpha}\lambda^2 d_2\! & -\sqrt{\alpha} B & 2\gamma \alpha\lambda d_3
\end{matrix}
\!\right) \!\left(\!
\begin{matrix}
\sqrt{\alpha} q_{-1}(t)
\\
q_0(t)
\\
p_{-1}(t)
\\
\sqrt{\alpha} p_0(t)
\end{matrix}\!
\right)\!,\!\!\!
\\
\label{n2}
\frac{d}{dt}\left(
\begin{matrix}
\vec{q}(t)
\\
\vec{p}(t)
\end{matrix}
\right)=2 \left(
\begin{matrix}
-\gamma \alpha\lambda d_3+C\vec{p}(t)\cdot \vec{q}(t) & \gamma \alpha\lambda d_1-C\vec{q}\,{}^2(t)
\\
-\gamma \alpha\lambda d_2+C\vec{p}\,{}^2(t) & \gamma \alpha\lambda d_3-C\vec{p}(t)\cdot \vec{q}(t)
\end{matrix}
\right) \left(
\begin{matrix}
\vec{q}(t)
\\
\vec{p}(t)
\end{matrix}
\right),
\end{gather}
where
\begin{gather*}
B:=2(\lambda -\epsilon)a\left(\gamma-\nu(\lambda-\epsilon)\vec{\mu}^2\right),
\qquad
C:=(\lambda-\epsilon)\left(\gamma\alpha-\nu(\lambda-\epsilon)a^2\right).
\end{gather*}
Integrating linear system given in~\eqref{nn1} we obtain
\begin{gather}
\label{m4}
\left(
\begin{matrix}
\sqrt{\alpha} q_{-1}(t)
\\
q_0(t)
\\
p_{-1}(t)
\\
\sqrt{\alpha} p_0(t)
\end{matrix}
\right)= \left(
\begin{matrix}
D(t) & E(t) & F(t) & G(t)
\\
-E(t) & D(t) & -G(t) & F(t)
\\
I(t) & J(t) & K(t) & L(t)
\\
-J(t) & I(t) & -L(t) & K(t)
\end{matrix}
\right) \left(
\begin{matrix}
\sqrt{\alpha} q_{-1}(0)
\\
q_0(0)
\\
p_{-1}(0)
\\
\sqrt{\alpha} p_0(0)
\end{matrix}
\right),
\end{gather}
where
\begin{gather*}
D(t)=\left(-\frac{\alpha\lambda}{\sqrt{-\delta_{\lambda,\alpha}}} d_3 \sinh (2\gamma
\sqrt{-\delta_{\lambda,\alpha}}t)+\cosh (2\gamma \sqrt{-\delta_{\lambda,\alpha}}t) \right) \cos(\sqrt{\alpha}Bt),
\\
E(t)=\left(-\frac{\alpha\lambda}{\sqrt{-\delta_{\lambda,\alpha}}} d_3 \sinh (2\gamma
\sqrt{-\delta_{\lambda,\alpha}}t)+\cosh (2\gamma \sqrt{-\delta_{\lambda,\alpha}}t) \right) \sin(\sqrt{\alpha}Bt),
\\
F(t)=\frac{d_1\sqrt{\alpha}}{\sqrt{- \delta_{\lambda,\alpha}}} \sinh (2\gamma \sqrt{-\delta_{\lambda,\alpha}}t)
\cos(\sqrt{\alpha}Bt),
\\
G(t)=\frac{d_1\sqrt{\alpha}}{\sqrt{- \delta_{\lambda,\alpha}}} \sinh (2\gamma \sqrt{-\delta_{\lambda,\alpha}}t)
\sin(\sqrt{\alpha}Bt),
\\
I(t)=\frac{-\alpha \sqrt{\alpha}\lambda^2 d_2}{\sqrt{- \delta_{\lambda,\alpha}}} \sinh (2\gamma
\sqrt{-\delta_{\lambda,\alpha}}t) \cos(\sqrt{\alpha}Bt),
\\
J(t)=\frac{-\alpha \sqrt{\alpha}\lambda^2 d_2}{\sqrt{-\delta_{\lambda,\alpha}}} \sinh (2\gamma
\sqrt{-\delta_{\lambda,\alpha}}t) \sin(\sqrt{\alpha}Bt),
\\
K(t)=\left(\frac{\alpha\lambda}{\sqrt{-\delta_{\lambda,\alpha}}} d_3 \sinh (2\gamma \sqrt{-\delta_{\lambda,\alpha}}t)
+\cosh (2\gamma \sqrt{-\delta_{\lambda,\alpha}}t) \right) \cos(\sqrt{\alpha}Bt),
\\
L(t)=\left(\frac{\alpha\lambda}{\sqrt{-\delta_{\lambda,\alpha}}} d_3 \sinh (2\gamma \sqrt{-\delta_{\lambda,\alpha}}t)
+\cosh (2\gamma \sqrt{\delta_{\lambda,\alpha}}t) \right) \sin(\sqrt{\alpha}Bt).
\end{gather*}
Further, substituting solutions~\eqref{m4} into~\eqref{n2}, see also~\eqref{ee11}--\eqref{ee13}, we come to
a~non-auto\-no\-mous linear system of equations for functions $\vec{q}(t)$ and $\vec{p}(t)$.
In order to solve this system let us consider separately two subcases $a \neq 0$ and $a=0$.

If $a\neq 0$ then from~\eqref{e3a1} we get
\begin{gather}
\label{n3}
\left(
\begin{matrix}
\vec{q}(t)
\\
\vec{p}(t)
\end{matrix}
\right)=\frac{1}{\lambda a} \left(
\begin{matrix}
\lambda q_0(t) & -\alpha\lambda q_{-1}(t)
\\
p_0(t) & -p_{-1}(t)
\end{matrix}
\right) \left(
\begin{matrix}
\vec{x}(t)
\\
\vec{y}(t)
\end{matrix}
\right),
\end{gather}
where $q_{-1}(t)$, $q_{0}(t)$, $p_{-1}(t)$ and $p_0(t)$ are given by~\eqref{m4} and $(\vec{x}(t),\vec{y}(t))$ were found
in Section~\ref{Section2}, see~\eqref{m5}, \eqref{m6}.

In the case $a=0$ one has $\vec{x}(t)\times \vec{y}(t)=0$.
So, instead of~\eqref{n3} we consider the equations
\begin{gather}
\label{m7}
p_0(t)\vec{\mu}=\vec{p}(t)\times \vec{y}(t),
\qquad
\alpha\lambda q_{-1}(t)\vec{\mu}=\vec{q}(t)\times \vec{x}(t),
\end{gather}
which also follows from~\eqref{e3a1}.
From~\eqref{e3a1} and~\eqref{ee11}, \eqref{ee12} we have
\begin{gather}
\vec{\mu}\cdot \vec{q}(t)=0,
\qquad
\vec{\mu}\cdot \vec{p}(t)=0,
\nonumber
\\
\vec{q}\,{}^2(t)=d_1-\alpha\lambda q_{-1}^2(t)- \lambda q_0^2(t),
\qquad
\vec{p}\,{}^2(t)=d_2-\frac{1}{\alpha\lambda}p_{-1}^2(t)- \frac{1}{\lambda} p_0^2(t).
\label{l8c}
\end{gather}
The functions $\vec{x}(t)$ and $\vec{y}(t)$ we f\/ind solving equations~\eqref{a52a2},~\eqref{a52a3} which in considered case
reduce to the linear system
\begin{gather}
\frac{d\vec{x}}{dt}=2(\epsilon -\lambda)\gamma \alpha\vec{\mu}\times \vec{x},
\qquad
\label{m2}
\frac{d\vec{y}}{dt}=2(\epsilon -\lambda)\gamma \alpha\vec{\mu}\times \vec{y}.
\end{gather}
Solution of~\eqref{m2} is given by
\begin{gather*}
\vec{x}(t)=O_{\vec{\mu}}(t)\vec{x}(0)
\qquad
\text{and}
\qquad
\vec{y}_{\vec{\mu}}(t)=O_{\vec{\mu}}(t)\vec{y}(0),
\end{gather*}
where $O_{\vec{\mu}}(t)\in {\rm SO}(3)$ is the rotation on the angle $2(\epsilon -\lambda)\gamma \alpha t$ around the constant
angular momentum vector $\vec{\mu}$.
Now, assuming $\mu_1=\mu_2=0$ after solving algebraic system of equations given by~\eqref{m7}, \eqref{l8c} we easily f\/ind
$\vec{q}(t)$ and $\vec{p}(t)$.

Finally let us discuss a~few possible physical interpretations of the above integrated Hamiltonian systems.

Firstly let us note that if $\gamma=1$ and $\epsilon=\lambda$ then $h=\frac{1}{\alpha\lambda}c_1\circ\iota^{-1}\circ{\cal J}=\delta_{\lambda,\alpha}$.
In this case equations~\eqref{r1} take the form
\begin{gather}
\label{1r}
\frac{d}{d t} \left(
\begin{matrix}
\eta_{\lambda,\alpha}q
\\
p
\end{matrix}
\right) =-2\left(
\begin{matrix}
d_3 {\bf 1_5} & -d_1 {\bf 1_5}
\\
d_2 {\bf 1_5} & -d_3 {\bf 1_5}
\\
\end{matrix}
\right) \left(
\begin{matrix}
\eta_{\lambda,\alpha}q
\\
p
\end{matrix}
\right).
\end{gather}
Since
\begin{gather*}
\{h, \delta_{\lambda, \alpha}\}=0
\qquad
\text{and}
\qquad
\{h,\vec{d}\}=0
\end{gather*}
solution of~\eqref{1r} is given by
\begin{gather}
\label{n1}
\left(
\begin{matrix}
\eta_{\lambda,\alpha}q(t)
\\
p(t)
\end{matrix}
\right)= \Psi\left(A_{\lambda, \alpha}(t)\right) \left(
\begin{matrix}
\eta_{\lambda,\alpha}q(0)
\\
p(0)
\end{matrix}
\right),
\end{gather}
where
\begin{gather}
\label{s6}
A_{\lambda, \alpha}(t)= \exp \left({-}2 t \left(
\begin{matrix}
d_3 {\bf 1_5} & -d_1 {\bf 1_5}
\\
d_2 {\bf 1_5} & -d_3 {\bf 1_5}
\\
\end{matrix}
\right)\right)
\end{gather}
is a~one-parameter subgroup of ${\rm SL}(2,\mathbb{R})_{\vec{d}}$.
This allows us to restrict the Hamiltonian $\delta_{\lambda, \alpha}$ and the f\/low $\Psi (A_{\lambda,
\alpha}(t) )$ to symplectic submanifold of $T^{*}\mathbb{R}^5$ def\/ined by the equations $d_1=\const$ and $d_3=0$.
Such a~submanifold is the bundle $T^*Q_{\lambda,\alpha}$ cotangent to the quadric
$Q_{\lambda,\alpha}:=\{q\in\mathbb{R}^5$: $\alpha\lambda q_{-1}^2+\lambda q_0^2+\vec{q}\,{}^2=d_1=\const\}$.

The Hamiltonian $\delta_{\lambda, \alpha}$ after restriction to $T^*Q_{\lambda,\alpha}$ represents kinetic energy
\begin{gather}
\delta_{\lambda, \alpha}= d_1d_2=d_1\left(\frac{1}{\alpha\lambda} p_{-1}^2+\frac{1}{\lambda}
p_{0}^2+\vec{p}\,{}^2\right)
=\frac 12 m \left(\frac{d q}{d t}\right)^{\top}\eta_{\lambda,\alpha} \left(\frac{d q}{dt}\right)
\nonumber
\\
\phantom{\delta_{\lambda, \alpha}}
= d_1\left(\alpha\lambda \left(\frac{d q_{-1}}{d t}\right)^2+\lambda \left(\frac{d q_{0}}{d t}\right)^2+\left(\frac{d
\vec{q}}{d t}\right)^2\right)
\label{2r}
\end{gather}
of the free particle localized on the quadric $Q_{\lambda,\alpha}$.
In~\eqref{2r} we identify $2 d_1$ with the mass of the particle and express momentum~$p$ by velocity $\frac{d q}{d t}$
by means of metric tensor
\begin{gather*}
p=\eta_{\lambda,\alpha} \frac{d q}{d t}.
\end{gather*}
Therefore~\eqref{n1} is the geodesic f\/low on the four-dimensional hypersurface $Q_{\lambda,\alpha}$ which is for
example:
\begin{enumerate}[i)]\itemsep=0pt
\item $S^4$ if $\alpha=\lambda=1$,
\item de Sitter spaces ${\rm dS}_4$ if $\alpha=\lambda=-1$,
\item anti-de Sitter spaces ${\rm AdS}_4$ if $\alpha=1$ and $\lambda=-1$.
\end{enumerate}

Hamiltonian~\eqref{e4} generalizes dynamics generated by Hamiltonian~\eqref{2r} in two aspects.
Firstly, it contains interaction counterparts of the free energy Hamiltonian~\eqref{2r}.
Secondly, one can reduce the system~\eqref{r1}, \eqref{e4} to various invariant submanifolds of $T^*\mathbb{R}^5$.
In particular, after reducing it to symplectic manifold, which is mapped by the momentum map ${\cal J}$ on the coadjoint
orbit ${\cal J}(({\cal I}^{-1}(\vec{d})\cap T_{\reg}^{*}\mathbb{R}^5) /{\rm SL}(2,\mathbb{R})_{\vec{d}}
)$, we come back to the system~\eqref{a52}--\eqref{a52a3} restricted to this coadjoint orbit.
Let us recall here that Hamiltonian f\/low $\sigma_{t}^{h}$ def\/ined by Hamiltonian~\eqref{e4} commutes with the action of
${\rm SL}(2,\mathbb{R})_{\vec{d}}$ and $\vec{d}$ is an integral of motion for this f\/low.
Since for $\vec{d}\in\Delta_s$ one has ${\cal J}(({\cal I}^{-1}(\vec{d})\cap T_{\reg}^{*}\mathbb{R}^5)
/{\rm SL}(2,\mathbb{R})_{\vec{d}} )\subset \Omega_s$ we can consider $(\vec{x},\vec{y})$ as a~local coordinates on
$({\cal I}^{-1}(\vec{d})\cap T_{\reg}^{*}\mathbb{R}^5) /{\rm SL}(2,\mathbb{R})_{\vec{d}}$.
The above follows from~\eqref{x2x} and~\eqref{ab55}.
Restricting integrals of motion $I_1$, $I_2$, $I_3$ and $I_4$ def\/ined by~\eqref{l1c}
to $({\cal I}^{-1}(\vec{d})\cap T_{\reg}^{*}\mathbb{R}^5) /{\rm SL}(2,\mathbb{R})_{\vec{d}}$ we f\/ind three integrals of motion on
$({\cal I}^{-1}(\vec{d})\cap T_{\reg}^{*}\mathbb{R}^5) /{\rm SL}(2,\mathbb{R})_{\vec{d}}$.
\begin{gather*}
\tilde{I}_1:=\lambda \left(I_1\circ {\cal J}\right)\left(I_2\circ {\cal J}\right)=\left(\vec{x}\times\vec{y}\right)_3=x_1y_2-x_2y_1,
\\
\tilde{I}_2:=c_1+\dfrac{\lambda}{\lambda-\epsilon} I_3\circ {\cal J}=\vec{x}\,{}^2+\alpha \vec{y}\,{}^2,
\qquad
\tilde{I}_3:= I_4\circ {\cal J}=\left(\dfrac{\lambda-\epsilon}{\lambda}\right)^2 \left(\vec{x}\times \vec{y}\right)^2
\end{gather*}
being in involution.
The integral of motion $I_1\circ {\cal J}=a$, as it follows from equation
\begin{gather*}
\lambda^2a^4+(\tilde{I}_2-c_1)a^2+\dfrac{\alpha\lambda}{(\lambda-\epsilon)^2}\tilde{I}_3=0,
\end{gather*}
is functionally dependent on $\tilde{I}_2$ and $\tilde{I}_3$.
The rank of the $6\times 3$ Jacobi matrix $D\tilde{I}(\vec{x},\vec{y})$ of the map
$\tilde{I}:\mathbb{R}^6\longrightarrow \mathbb{R}$ is equal three if\/f $(\vec{x},\vec{y})\in \mathbb{R}^6\setminus \Sigma
$, where the closed subset $\Sigma\subset \mathbb{R}^6$ is def\/ined as the intersection of zero levels of all $3\times 3$
minors of $D\tilde{I}(\vec{x},\vec{y})$.
Thus we conclude that $\tilde{I}_2$, $\tilde{I}_2$ and $\tilde{I}_3$ are functionally independent almost everywhere on
$\big({\cal I}^{-1}(\vec{d})\cap T_{\reg}^{*}\mathbb{R}^5\big) /{\rm SL}(2,\mathbb{R})_{\vec{d}}$.

In order to obtain some other interpretation of Hamiltonian systems integrated above let us reduce the canonical
one-form
\begin{gather*}
\gamma =p_{-1}\mathrm{d}q_{-1}+p_0 \mathrm{d}q_0+\vec{p}\cdot \mathrm{d}\vec{q}
\end{gather*}
of $T^*\mathbb{R}^5$ to $T^*Q_{\lambda,\alpha}$.
From $d_1=\const$ and $d_3=0$ we f\/ind that
\begin{gather*}
q_{-1}=\pm \sqrt{\frac{1}{\alpha\lambda}\big(d_1-\lambda q_0^2-\vec{q}\,{}^2\big)},
\qquad
p_{-1}=\frac{-1}{q_{-1}} (q_0p_0 +\vec{q}\cdot \vec{p} )
\end{gather*}
and thus
\begin{gather*}
\gamma\big|_{T^*Q_{\lambda,\alpha}}=\pi_0 \mathrm{d} q_0+ \vec{\pi}\cdot \mathrm{d}\vec{q},
\end{gather*}
where
\begin{gather*}
\pi_0=p_0+ \lambda \dfrac{q_0p_0 +\vec{q}\cdot \vec{p}}{d_1-\lambda q_0^2-\vec{q}\,{}^2}q_0,
\qquad
\vec{\pi}=\vec{p}+ \dfrac{q_0p_0 +\vec{q}\cdot \vec{p}}{d_1-\lambda q_0^2-\vec{q}\,{}^2}\vec{q}.
\end{gather*}
For $\alpha\lambda=1$ Hamiltonian~\eqref{e4}, after reduction to $T^*Q_{\lambda,\alpha}$, takes in the canonical
coordinates ($q_0, \vec{q},\pi_0, \vec{\pi}$) a~form of polynomial of degree eight
\begin{gather}
h=\gamma d_1\lambda^{-1}\pi_0^2+ \gamma\lambda^{-1}(\epsilon-\lambda)(\vec{\pi}\times \vec{q})^2
+\gamma d_1\vec{\pi}^2-\gamma  (\pi_0q_0+\vec{\pi}\cdot \vec{q} )^2
\nonumber
\\
\phantom{h=}
{}+(\lambda-\epsilon)\lambda^{-2} \big(\nu (\lambda-\epsilon) (\vec{\pi}\times \vec{q})^2 -\gamma\big)
\big(d_1-\lambda q_0^2-\vec{q}\,{}^2\big) \pi_0^2.
\label{p1}
\end{gather}
Passing in~\eqref{p1} to complex coordinates ($z_0=q_0+i\pi_0$, $\vec{z}=\vec{q}+i\vec{\pi}$) we obtain Hamiltonian
\begin{gather}
h= \frac 14\bigg(\gamma d_1 \big(2\vec{z}\cdot\bar{\vec{z}}-\vec{z}\,{}^2-\bar{\vec{z}}\,{}^2\big)+ \gamma d_1 \lambda^{-1}
\big(2|z_0|^2 -z_0^2-\bar{z}_0^2\big)
\nonumber
\\
\phantom{h=}
{}+\frac{\gamma (\lambda-\epsilon)}{\lambda} (\vec{z}\times\bar{\vec{z}})^2 + \frac{\gamma}{16}
\big(z_0^2-\bar{z}_0^2+\vec{z}\,{}^2-\bar{\vec{z}}\,{}^2 \big)^2
\nonumber
\\
\phantom{h=}
{}+\frac{\lambda-\epsilon}{16\lambda^2} \big(\nu(\lambda-\epsilon)(\vec{z}\times\bar{\vec{z}})^2+4\gamma \big)
\big(z_0^2+\bar{z}_0^2-2|z_0|^2\big)
\nonumber
\\
\phantom{h=}
\times
\big(4d_1-\lambda z_0^2-\lambda \bar{z}_0^2 -2\lambda |z_0|^2- \vec{z}\,{}^2-\bar{\vec{z}}\,{}^2-2\vec{z}\cdot\bar{\vec{z}}\big)\bigg),
\label{p2}
\end{gather}
which describes a~system of four running plane waves, slowly varying in nonlinear dielectric medium.
The terms in~\eqref{p2} higher than quadratic ones are responsible for such nonlinear optical ef\/fects as
intensity-dependent phase shift (Kerr ef\/fect) and the conversion between the modes.
In a~similar way one can interpret Hamiltonian~\eqref{e4}, rewritten in complex coordinates, to describe a~system of
f\/ive nonlinear running plane waves.
We refer to~\cite{12} and~\cite{13} for the treatment of Hamiltonian formulation of propagation of optical traveling
wave pulses.
Also one can f\/ind this type of nonlinear Hamiltonian optical system integrated by quadratures in~\cite{14}.

\subsection*{Acknowledgements}

Authors are grateful to the f\/irst referee for invaluable remarks which allowed us to avoid mistakes and make the paper
more readable.

\pdfbookmark[1]{References}{ref}
\LastPageEnding

\end{document}